\documentclass[useAMS,usenatbib,referee]{mn2e}
\usepackage[dvips]{graphicx}
\usepackage{wrapfig,epsfig,epsf}
\usepackage{graphics}
\usepackage{amsmath}

\title[Relativistic twisted disc around a Kerr black hole]{A fully
relativistic twisted disc 
around a slowly rotating Kerr black hole: derivation of
dynamical equations and the shape of stationary configurations}
\author[V. V. Zhuravlev and P. B. Ivanov]{V.V. Zhuravlev$^{1}$\thanks{E-mail:
v.jouravlev@gmail.com (VZh)} and P.B. Ivanov$^{2}$\thanks{E-mail: pbi20@cam.ac.uk (PBI)}\\
$^{1}$Sternberg Astronomical Institute, Moscow State University, Universitetskij pr., 13, 119992 Moscow, Russia\\
$^{2}$  Astro Space Centre, P.N. Lebedev Physical Institute, 84/32
Profsoyuznaya Street, Moscow, 117810, Russia}

\begin{document}

\date{Accepted. Received; in original form}

\pagerange{\pageref{firstpage}--\pageref{lastpage}} \pubyear{2010}

\maketitle

\label{firstpage}

\begin{abstract}
In this paper we derive equations describing dynamics and
stationary configurations of a twisted fully relativistic thin
accretion disc around a slowly rotating black hole. We assume that
the inclination angle of the disc is small and that the standard
relativistic generalisation of the $\alpha $ model of accretion
discs is valid when the disc is flat. We find that similarly to
the case of non-relativistic twisted discs the disc dynamics and
stationary shapes can be determined by a pair of equations
formulated for two complex variables describing orientation of the
disc rings and velocity perturbations induced by the twist.

We analyse analytically and numerically the shapes of stationary
twisted configurations of accretion discs having non-zero
inclinations with respect to the black hole equatorial plane at
large distances $r$ from the black hole. It is shown that the
stationary configurations
depend on two parameters - the viscosity parameter $\alpha $ and
the parameter $\tilde \delta = \delta_{*}/\sqrt a$, where
$\delta_{*}$ is the opening angle ($\delta_{*}\sim h/r$, where $h$
is the disc half thickness and $r$ is large) of a flat disc and
$a$ is the black hole rotational parameter. When $a > 0$ and
$\tilde \delta \ll 1$ the shapes depend drastically on value of
$\alpha$. When $\alpha $ is small the disc inclination angle
oscillates with radius with amplitude and radial frequency of the
oscillations dramatically increasing towards the last stable
orbit, $R_{ms}$. When $\alpha $ has a moderately small value the
oscillations do not take place but the disc does not align with
the equatorial plane at small radii. The disc inclination angle is
either increasing towards $R_{ms}$ or exhibits a non-monotonic
dependence on the radial coordinate. Finally, when $\alpha $ is
sufficiently large the disc aligns with the equatorial plane at
small radii. When $a < 0$ the disc aligns with the equatorial
plane for all values of $\alpha $.

The results reported here may have implications for determining
structure and variability of accretion discs close to $R_{ms}$ as
well as for modelling of emission spectra coming from different
sources, which are supposed to contain black holes.

\end{abstract}


\section{Introduction}

Twisted accretion discs around rotating black holes are used for
explanations of different observational phenomena. For example
they are invoked to explain jet precession observed in certain
active galactic nuclei (AGN) (e.g. \citet{Cap04} and references
therein ), warping of the accretion disc in the maser galaxy NGC
4258 (e.g. \citet{PTL98,Cap07,M08}), certain features of emission
line profiles coming from AGN (e.g. \citet{B99,Cad03,WCY10}) etc..
\citet{FO08,FO09} proposed that shear velocities induced by the
disc twist (see below) provide  a mechanism of excitation of high
frequency quasi-periodic oscillations observed in many
astronomical objects. Light curves, forms of emission lines of
precessing twisted tori around a Kerr black hole as well as their
shapes as seen from large distances have recently been
investigated numerically by \citet{DF10}.

Theoretical studies of thin twisted discs started from the seminal
paper \citet{BP75}, who proposed a simple model of a stationary
twisted disc, where a twisted disc was considered as a collection
of circular rings interacting with each other due to viscous
forces. In frameworks of this model Bardeen and Petterson came to
the conclusion that a stationary twisted disc inclined with respect to
the black hole equatorial plane at large distances from the black
hole aligns with this plane at smaller radii - the effect called
later as "the Bardeen-Petterson effect". This alignment was
explained as being due to the presence of 'gravitomagnetic force'
acting from the side of a rotating black hole on the disc rings.
This force causes precession of orbital planes of free particles
around a direction of the black hole spin and breaks the spherical
symmetry of the problem. The Bardeen and Petterson approach was
later developed and generalised on non-stationary twisted discs by
e.g. \citet{P77,P78,Hat81}.

In frameworks of hydrodynamical theory of perturbations \citet{PP83}
pointed out that the simple model developed in the previous works is, in fact,
inconsistent since it does not conserve all components of the angular
momentum content of an accretion disc. They showed that a
self-consistent model must, necessarily, contain perturbations of
the disc density and velocity fields induced by the disc twist,
which determine deviations of the gas particles motion in the disc
from the circular one (see \citet{II97} for a
qualitative explanation of this effect) . In particular, the
velocity perturbations should be odd with respect to the disc
vertical coordinate thus having the structure of a shear flow. \citet{PP83}
also found that when a sufficiently viscous disc is
considered in the classical Newtonian gravitational field of a
point mass a typical alignment scale of a stationary twisted
configuration decreases with decrease of the \citet{SS73}
$\alpha $ parameter (see also e.g. \citet{KP85})
and that non-stationary propagation of twisted disturbances
through the disc has a diffusive character with a characteristic
time scale being also proportional to $\alpha $ (also
e.g. \citet{Kum90}) contrary to what was claimed in the previous
studies. These two effects are determined by the well known
degeneracy of the Keplerian potential, which has orbital and
epicyclic frequencies equal to each other. It was later shown by
\citet{PL95} that in the opposite case of a low
viscosity Keplerian disc propagation of non-stationary twisted
disturbances through the disc has a wave-like character with a
typical speed of the order of speed of sound in the disc. These
studies considered the gravitational field of a black hole as a
field of a Newtonian point mass with the gravitomagnetic force
being treated as a classical force causing precession of the disc
rings.

\citet{II97} considered post-Newtonian corrections to equations
describing velocity perturbations. They found that when a low
viscosity stationary twisted disc is considered and the disc gas
is rotating in the same sense as a black hole the
Bardeen-Petterson effect does not take place. Instead, there
are radial oscillations of the disc inclination angle with
amplitude increasing with decrease of the distance from the black
hole. This effect was later confirmed by time dependent
calculations made by \citet{LOP02}. \citet{II97} also provided a
qualitative explanation of this effect and a general criterion of
appearance of these oscillations. Namely, they take place in a low
viscosity stationary twisted disc when the signs of the apsidal
and nodal precessions are the same. In the opposite case the disc
aligns with the equatorial plane of the black hole for any value
of viscosity. For example, the latter situation is realised when
the black hole rotates in the direction opposite to the direction
of the orbital motion of the gas in the disc, see \citet{II97} and below
or when a Newtonian twisted disc around a massive binary star or a
binary black hole is considered, e.g.  \citet{ipp}. We are going to show below
that this criterion is applicable to our fully relativistic problem as well
and that whether the disc aligns with the equatorial plane or not depends on the
direction of rotation of the black hole relative to the orbital motion.

Additionally, \citet{II97,DI97} \footnote{Note that there is a
number of misprints in \citet{DI97}. The final equations of this
paper are, however, correct} showed that the linear analysis made
by \citet{PP83} and \citet{PL95} can be extended to the case of
disc inclination angles, which are larger than the disc opening
angle $\delta_{*}\approx h/r\ll 1$, where $h$ is the disc half
thickness and $r$ is the radial coordinate, and the precise
definition of the angle $\delta_{*}$ is made in such way that it
is a constant independent on $r$, see equations (\ref{eq31}) and
(\ref{eq32}) below. This was done by employing a formalism based
on the so-called "twisted coordinate" system introduced by
\citet{P77,P78}. An analogous coordinate system has later been
used by \citet{Og99} to consider the disc inclination angles of
order of unity and, accordingly,  to construct a non-linear theory
of evolution of sufficiently viscous twisted discs.

As we described above, all previous formalisms of description of the
thin twisted discs were based either on the classical or
post-Newtonian treatments of the problem\footnote{Note, however,
that dynamical models of fully relativistic accretion tori have
been recently considered by numerical means, e.g.
\citet{FB08,DF10}.}. It is, however, important to consider the
problem in full General Relativity. This is especially crucial
either for the discs having a small value of $\alpha \ll
\delta_{*}$ or in the case of slowly rotating black holes having
their rotational parameter $a \ll 1$, where a stationary accretion
disc can be twisted at scales comparable to the black hole
gravitational radius, $r_g$. A formalism based on General
Relativity can also be very useful for a quantitative modelling of
emission spectra coming from twisted disc as well as for accurate
studying of many other physical processes occurring in such discs
like e.g. their self-irradiation.

In this paper we construct a fully relativistic theory of thin
twisted discs at radii larger than the radius of the last 
stable orbit around slowly rotating black holes, treating the
effect of the black hole rotation in the linear approximation
only. This essentially means that only one term in general
equations of motion (proportional to the rotational parameter $a$
and describing the gravitomagnetic force) can be taken into account
and all quantities entering in our final equations describing
dynamics of twisted discs can be defined with help of the
Schwarzschild metric of a non-rotating black hole. We make the
usual assumptions of smallness of the disc opening angle
$\delta_{*}$, the disc inclination angle $\beta $ with respect to
the equatorial plane and its logarithmic derivative, $r{d\over
dr}\beta \ll 1$. Additionally, we neglect self-gravity of the
discs. We generalise the twisted coordinate system of
\citet{P77,P78} to the relativistic case and write down our
general equations of motion in this generalised coordinate system.
We show that these equations can be naturally split on two
different subsets: 1) a set of equations describing the standard
flat disc accretion and 2) a set of equations determining dynamics
of variables related to the disc twist. As a set of the standard
equations we use equations of the well known \citet{NT73}
accretion disc model, which is the relativistic generalisation of
the \citet{SS73} $\alpha $-disc model. Equations of the set 2) are
used to derive a pair of equations (\ref{eq50}) and (\ref{eq51})
fully describing the dynamics and stationary configurations of
twisted discs under our approximations. Similar to the case of
Newtonian twisted discs (see e.g. \citet{DI97}) these equations
can be formulated for two dependent complex variables - ${\bf
W}(r,t)=\beta e^{\gamma }$, where $\gamma $ is the second Euler
angle determining longitude of ascending nodes of the disc rings,
and ${\bf B}(r,t)$ determining the form of velocity perturbations
in the disc.

In the second part of the paper we study in detail shapes of
stationary configurations, which are described by equation
(\ref{eq52}). We show that this equation contains only two
independent parameters - $\alpha $ assumed to be a constant
throughout the disc and $\tilde \delta =\delta_{*}/\sqrt a$. When
the disc gas rotates in the same direction as the black hole ($a
>0 $), for the gas pressure dominated \citet{NT73}
models we find that there are three possible qualitatively
different shapes of the twisted discs depending on values of
$\alpha $. When $\alpha $ is very small the disc exhibits
oscillations of the inclination angle, which have a different
character of amplitude and radial frequency behaviour at large distances from the black
hole and close to the last stable orbit. The disc behaviour at
large radii is the same as described by \citet{II97}. On the other hand oscillations
of the inclination angle
close to the last stable orbit proceed in a different regime, with
the amplitude and the radial frequency strongly
growing towards the last stable orbit. When $\alpha $ has
moderately small values the oscillations are suppressed but the
disc does not align with the black hole equatorial plane. The
inclination angle is either monotonically growing towards the last
stable orbit or exhibits a non-monotonic  behaviour decreasing
with decrease of $r$ at large radii and increasing again towards
the last stable orbit at smaller radii. Finally, when $\alpha $ is
sufficiently large the Bardeen-Petterson effect takes place.

In Section 2 we present our basic definitions, introduce the
twisted coordinate system, describe fundamental equations of
motions and the flat disc model used in our study. In Section 3 we
derive our final equations (\ref{eq50}) and (\ref{eq51})
describing the dynamics and stationary configuration of the
twisted discs. Section 4 is devoted to analytical and numerical
studies of the stationary twisted configurations. Note that
Section 3 and Section 4 can be read independently.

We use the natural systems of units throughout the paper setting
the gravity constant and the speed of light to unity. The metric
signature is $(+,-,-,-)$ and the usual summation rule over
repeating indices is implied.

\section{Basic definitions and equations}

\subsection{Metric}

As was mentioned in Introduction we assume that the black hole
rotates slowly with the rotational parameter $a \ll 1$ and derive
our dynamical equations taking into account only leading terms in
$a$. Therefore, for our purposes, it suffices to consider the
metric of a Kerr black hole in the Boyer-Lindquist coordinates taking
into account only linear in $a$ terms
\begin{equation}
ds^2 =  (1-2M/ R ) dt^2 -(1-2M/R)^{-1} dR^2 - R^2(d\theta^2 +
sin^2\theta d\phi^2) + 4a \frac{M}{R} sin^2\theta\, d\phi\, dt.
\label{eq1}
\end{equation}
It is easy to see that apart from the $g_{0\varphi}$ term the
metric coincides with the metric of a non-rotating black hole
written in the  Schwarzschild coordinates, see e.g. \cite{ZN}.

In the non-relativistic problems the dynamical equations for the
twisted discs take simplest form in the so-called "twisted
coordinates" introduced by \citet{P77,P78}. As we discuss in
Section 2.2 these coordinates are obtained from the cylindrical
ones by rotation. In order to find a simple relativistic
generalisation of the twisted coordinates we would like to bring
the metric (\ref{eq1}) into another, the so-called 'spatially
isotropic' form, where the spatial part of the line element is
proportional to the Cartesian line element. This is done by the
change of the radial coordinate
\begin{equation}
R = R_{I}\left (1+\frac{M}{2R_{I}} \right )^2 \label{eq2},
\end{equation}
which brings (\ref{eq1}) in the form
\begin{equation}
ds^2 = \left ( \frac{1-\frac{1}{2R_{I}}}{1+\frac{1}{2R_{I}}}
\right )^2 dt^2 - \left (1+\frac{1}{2R_{I}} \right )^4 (dR_{I}^2 +
R_{I}^2d\theta^2 + R_{I}^2 sin^2\theta d\phi^2) +
4\frac{a\,sin^2\theta}{R_{I} \left ( 1 +\frac{1}{2R_{I}} \right
)^2} dt\, d\phi \label{eq3},
\end{equation}
where the radial and time coordinates are expressed in units of
$M$ from now on. Introducing the cylindrical coordinates
$\{r=R_{I}\sin \theta,\phi, z=R_{I}\cos \theta\}$ we can rewrite
(\ref{eq3}) in an equivalent form
\begin{equation}
ds^2 = K_1^2 dt^2 + 2a r^2 K_1 K_3 d\phi dt - K_2^2(dr^2 + dz^2
+ r^2 d\phi^2) \label{eq4},
\end{equation}
where
\begin{equation}
K_1 = \frac{1-\frac{1}{2R_I}}{1+\frac{1}{2R_I}}, \quad K_2 = \left
( 1+\frac{1}{2R_I} \right )^2, \quad  K_3 =
\frac{2}{R_I^3}\frac{1}{1-\left (\frac{1}{2R_I} \right )^2}.
\label{eq5}
\end{equation}

The metric (\ref{eq3}) induces an associated orthonormal tetrad
of one-forms
\begin{equation}
\mbox{\boldmath$\omega$}^t = K_1 dt + a r^2 K_3 d\phi,
\quad \mbox{\boldmath$\omega$}^r = K_2 dr, \quad
\mbox{\boldmath$\omega$}^\phi = rK_2 d\phi, \quad
\mbox{\boldmath$\omega$}^z = K_2 dz. \label{eq6}
\end{equation}

We also use Cartesian coordinates $(x,y,z)$ related to $(r, \phi,
z)$ in the usual way.

\subsection{The twisted coordinate system}

The twisted coordinates $(\tau, r_{tw}, \psi, \xi)$ are obtained
from the Cartesian ones by rotation
\begin{equation}
\begin{array}{cccc}

\left ( \begin{array}{c} \tau \\ r_{tw} cos\psi \\ r_{tw} sin\psi \\
\xi
\end{array} \right ) & = &
\left ( \begin{array}{cccc}
1 & 0 & 0 & 0 \\
0 &  cos\gamma & sin\gamma & 0 \\
0 & -sin\gamma & cos\gamma & \beta  \\
0 & \beta sin\gamma & -\beta cos \gamma & 1
\end{array} \right )
& \left ( \begin{array}{c} t \\ x \\ y \\ z
\end{array} \right )
\end{array}, \label{eq7}
\end{equation}
where the Euler angles $\beta(\tau,r)$ and $\gamma(\tau,r)$ are
functions to be determined. The inclination angle $\beta $ is
assumed to be small. It is evident that $r_{tw}^2+\xi^2=r^2+z^2$.

Since only $r_{tw}$ is used below we omit the index $(tw)$ later
on. It is also convenient to use  $\varphi = \psi + \gamma$ in our
expressions and introduce the variables \begin{equation} \Psi_1 =
\beta \cos \gamma, \,\, \Psi_2 = \beta \sin \gamma \label{eq8}
\end{equation}
instead of $\beta $ and $\gamma $ and the quantities
\begin{equation} Z = \beta \sin \psi = \Psi_1 \sin \varphi -
\Psi_2 \cos \varphi, \quad  U=\dot Z, \quad W=
Z^\prime,\label{eqn8}
\end{equation}
where partial derivatives over $\tau$ and $r$ are denoted by dot
and prime,  respectively.

The orthonormal one-forms associated with the twisted coordinate
system are given by the expressions
\begin{equation} \mbox{\boldmath$\omega$}^\tau = (K_1-a r \xi K_3
\partial_\varphi U) d\tau +
                      a\xi K_3 \partial_\varphi (Z - r W )d r + ar K_3 (r-\xi Z) d\varphi -
                       a r K_3 \partial_\varphi Z d\xi,
\label{eq9}\end{equation}
\begin{equation}
\mbox{\boldmath$\omega$}^r = -\xi K_2 U d\tau +
K_2(1-\xi W) dr, \label{eq10}\end{equation}
\begin{equation}
 \mbox{\boldmath$\omega$}^\varphi = -\xi K_2
\partial_\varphi U d\tau -
                        \xi K_2 \partial_\varphi W dr + r K_2
                        d\varphi,
\label{eq11}\end{equation}
\begin{equation}
\mbox{\boldmath$\omega$}^\xi = r K_2 U d\tau + r K_2 W
dr + K_2 d\xi, \label{eq12}
\end{equation}
see Appendix A for their relation to the coordinate forms
(\ref{eq6}).

When $\beta=\beta^\prime=\gamma^\prime=0$ the twisted coordinates
are reduced to cylindrical coordinates rotated with respect to the
coordinates introduced above by angle $\gamma $ and the forms
(\ref{eq9}-\ref{eq12}) coincide with the expressions (\ref{eq6})
\footnote{Note that when $a=0$ it is sufficient to have
$\beta^\prime=\gamma^\prime=0$ for (\ref{eq9}-\ref{eq12}) to
coincide with (\ref{eq6}). Clearly, this is due to the spherical
symmetry of the Schwarzschild space-time.}.

The adjoint basis vectors ${\bf e}_{i}$ are obtained from the
duality condition $\mbox{\boldmath$\omega$}^i({\bf e
}_j)=\delta^i_j$. Explicitly, we have
\begin{equation}
{\bf e}_\tau = \frac{1}{K_1} \left ( \partial_\tau + \xi U
\partial_r + \frac{\xi}{r} \partial_\varphi U \partial_\varphi - r
U \partial_\xi \right ), \label{eq13}
\end{equation}
\begin{equation}
{\bf e}_r = \frac{1}{K_2} \left ( -a\xi\frac{K_3}{K_1}
\partial_\varphi Z \partial_\tau + (1+\xi W) \partial_r +
\frac{\xi}{r} \partial_\varphi W \partial_\varphi - r W
\partial_\xi  \right ),
\label{eq14}
\end{equation}
\begin{equation}
 {\bf e}_\varphi = \frac{1}{K_2} \left ( -a\frac{K_3}{K_1} (r
- \xi Z) \partial_\tau - a\xi \frac{ K_3}{K_1} r U \partial_r +
\left ( \frac{1}{r} - a\xi \frac{K_3}{K_1} \partial_\varphi U
\right ) \partial_\varphi + ar \frac{K_3}{K_1} r U \partial_\xi
\right ),\label{eq15}
\end{equation}
\begin{equation}
{\bf e}_\xi = \frac{1}{K_2} \left (  ar\frac{K_3}{K_1}
\partial_\varphi Z  \partial_\tau + \partial_\xi \right ).
\label{eq16}
\end{equation}

In the limit of large $r$ the expressions (\ref{eq9}-\ref{eq16})
tend to the corresponding expressions calculated for a flat
space-time in \citet{P78}.

We project all our dynamical quantities and equations of motion
onto the basis (\ref{eq9}-\ref{eq16}). In order to calculate
covariant derivatives in this basis the connection coefficients,
$\Gamma_{ijk}$, should be used. Their explicit forms are given in
Appendix A.

For our purposes we need covariant derivative of a vector and
covariant divergence of a tensor projected onto the bases
(\ref{eq9}-\ref{eq16}):
\begin{equation}
A^{i}_{;j}={\bf e}_{j}(A^{i})+\Gamma^{i}_{kj}A^{k}, \quad
A^{ij}_{;j} = {\bf e}_j (A^{ij}) + \Gamma_{kj}^i A^{kj} +
\Gamma^j_{kj} A^{ik}, \label{eq17}
\end{equation}
where $(;)$ stands for covariant derivative. Note that the indices
of components of the tensor quantities as well as the ones of the
connection coefficients projected onto our orthonormal bases can
be raised and lowered with help of the Minkowski metric
$\eta_{ik}$.

\subsection{Equations of motion}

Our equations of motion follow from the law of mass conservation
\begin{equation}
(\rho U^{i})_{;i}=0, \label{eq18}
\end{equation}
where $\rho$ is the rest mass density, $U^{i}$ are components of
four velocity, and equality to zero of covariant derivative of the
stress energy tensor,
\begin{equation}
T^{ik}_{;k}=0. \label{eq19}
\end{equation}

We consider the stress energy tensor of a viscous radiative fluid
\begin{equation}
T^{ik}=(\epsilon
+p)U^{i}U^{k}-pg^{ik}+T_{\nu}^{ik}-U^{i}q^{k}-U^{k}q^{i},
\label{eq20}
\end{equation}
where $\epsilon $ and $p$ are the energy density and pressure,
respectively, $g^{ik}$ and $T_{\nu}^{ik}$ are the components of
the metric and viscosity tensors, $q^{i}$ are the components of
radiation flux. We have $\epsilon =\rho + \epsilon_{th}$, where
the thermal energy $\epsilon_{th}$ is assumed to be much smaller
than the rest energy $\rho$. The viscous stress tensor
$T^{ik}_{\nu}=2\eta \sigma^{ik}$, where $\eta $ is dynamical viscosity,
\begin{equation}
\sigma^{ik}= {1\over 2}(U^{i}_{;j}P^{jk}+U^{k}_{;j}P^{ji})-{1\over
3}U^{j}_{;j}P^{ik}\label{eq21}
\end{equation}
is the shear tensor and
\begin{equation}
P^{ik}=g^{ik}-U^{i}U^{k} \label{eq22}
\end{equation}
is the projection tensor. Note that
\begin{equation}
U_{i}\sigma^{ik}=0, \quad \sigma^{i}_{i}=0. \label{eqn22}
\end{equation}

Similar to the analysis of Newtonian twisted discs in the fully
relativistic case in the linear approximation in angle $\beta $ we
can divide our equations of motion (\ref{eq18}) and (\ref{eq19}) in
two parts having different symmetries with respect to the
coordinates $\xi$ and $\varphi$: 1) a part describing an
unperturbed flat disc and 2) a part describing the disc's
perturbations associated with the disc's twist and warp. The first
(background) part  follows from standard models of relativistic
flat disc accretion, where all dynamical quantities should be
projected onto the basis (\ref{eq13}-\ref{eq16})
with $Z$, $U$ and $W$ formally set to zero.
In this paper we would like to consider the
simplest possible case of a relativistic geometrically thin
optically thick stationary $\alpha-$disc as a background model,
and use, accordingly, the results of \citet{SS73}, \citet{NT73},
hereafter NT, \citet{PT74} and \citet{RH95}, hereafter RH, to
describe it. The perturbation part is treated in a way similar to
what is done in \citet{DI97}, hereafter DI, it is discussed in
Section 3 with some technical details relegated to Appendix B.

\subsection{The background model of a flat relativistic disc}

For our purposes it is sufficient to consider only a few basic
properties of the flat disc models. Additionally, since we
consider the case of a slowly rotating black hole in this Section
we neglect corrections due to a nonzero value of $a$ and discuss
only accretion discs around a Schwarzschild black hole, formally
setting $a=0$. However, certain basic expressions accounting for
linear in $a$ corrections are shown in Appendix B, for
completeness.

Equations describing the relativistic flat disc models can
themselves be split in two parts: the ones responsible for disc's
structure and behaviour determined by effects occurring on a
dynamical time scale and the ones determining the disc's structure
and behaviour related to a slow viscous time scale.

Let us discuss the first group of equations temporarily neglecting
viscous interaction between neighbouring rings of the disc. In this
approximation the disc's gas is orbiting in the azimuthal
direction with nearly geodesic four-velocity having only two
components: $U^{\varphi}$ and
$U^{\tau}=\sqrt{1+(U^{\varphi})^{2}}$. We use, accordingly, the
standard expressions for geodesic circular motion in the field of
a Schwarzschild black hole
\begin{equation}
U^\varphi = (R-3)^{-1/2}=\left ( r-2+\frac{1}{4r} \right )^{-1/2},
\quad U^\tau = \left ( \frac{R-2}{R-3} \right
)^{1/2}=K_1K_2^{1/2}r^{1/2}U^{\varphi}, \label{eq23}
\end{equation}
where we set $R=K_{2}(r)r$ using the fact that we consider the
disc situated close to the equatorial plane of a black hole and
assume that the disc's gas is orbiting in positive direction with
respect to the angle $\varphi $. An important quantity associated
with the circular motion is its frequency with respect to a
distant observer, $\Omega \equiv {d\varphi \over d\tau}$. From
equation (\ref{eq23}) we get
\begin{equation}
\Omega = \frac{K_1}{K_2} \frac{U^\varphi}{r U^\tau}=R^{-3/2}.
\label{eq24}
\end{equation}

The disc structure in the vertical direction is governed by
equation of hydrostatic balance
\begin{equation}
{\partial_{\xi}p \over \rho}=-\left ( {U^{\varphi} \over r} \right )^{2}\xi,
\label{eq25}
\end{equation}
where we neglect $\epsilon_{th}\ll \rho$.

Viscous interactions in the disc result in a slow drift of the
disc's gas in the radial direction, and, accordingly, an
additional radial component of four velocity, $U^{r}$, appears.
Equations determining disc's properties related to this effect
follow from the laws of conservation of the rest mass, energy and
angular momentum, radiation transfer in the vertical direction and
properties of viscous interactions in the disc. They allow one to
complete the set of equations of the disc structure and obtain
explicit expressions for basic quantities describing the disc. Let
us discuss some of these equations relevant for our purposes.

The law of the rest mass conservation may be written in the form
\begin{equation}
-{\dot M\over 2\pi}=\Sigma K_{1}K_{2}^{2}rU^{r}, \label{eq26}
\end{equation}
where the surface density
\begin{equation}
\Sigma=\int d\xi \rho, \label{eq26n}
\end{equation}
and the rate of flow of the rest mass through the disc, $\dot M$,
is a constant.

In the flat disc's models only $(r\varphi)$ and $(r\tau)-$
components of the viscous stress tensor are important. As is shown
in NT and \citet{PT74} (see also RH) the laws of energy
and angular momentum conservation result in an explicit expression
for integrated over the vertical coordinate $(r\varphi)$-component
of the stress tensor \footnote{Note that this component is defined
in NT and RH with respect to another, nearly comoving, frame of
basis vectors. The component in our frame can be obtained
multiplying the NT and RH result by $U^{\tau}$. Also, integration
of this component in these papers is performed over the proper
vertical length $\xi_{proper}=K_{2}\xi$, see (\ref{eq4}).
Therefore, in order to obtain a vertically integrated value of
$T^{r\varphi}$ the NT and RH result should be additionally divided
by $K_{2}$.}
\begin{equation}
\bar  T^{r\varphi}={\dot M \over 2\pi}U^{\tau}r^{-3/2}{D(r)\over
K_2^{5/2}K_{1}^2},\label{eq26nn}
\end{equation}
where the bar stands hereafter for vertically integrated
quantities,
\begin{equation}
D=1-{\sqrt 6\over y}-{\sqrt 3\over 2y}\ln{{(y-\sqrt 3)(3+2\sqrt
2)\over (y+\sqrt 3)}}, \label{eq27}
\end{equation}
and $y=\sqrt {R}=\sqrt {K_{2} r}$. Note that $D=0$ at the
marginally stable orbit, when $R=R_{ms}=6$, where the disc
truncates. The value of $T^{r\tau}$ is obtained from
(\ref{eqn22}): $T^{r\tau}={U^{\varphi}\over
U^{\tau}}T^{r\varphi}$. From (\ref{eq21}) and (\ref{eq23}) it
follows another expression for $\bar T^{r\varphi}$,
\begin{equation}
\bar T^{r\varphi}={3\over 2}\bar \eta {K_1\over
K_{2}}(U^{\tau})^{2}U^{\varphi}/r.
 \label{eq28}
\end{equation}
Equating (\ref{eq28}) to (\ref{eq26nn}) we get
\begin{equation}
\bar \eta ={\dot M \over 3\pi}({r^{-1/2}\over
U^{\tau}U^{\varphi}}{D\over K_1^3K_2^{3/2}}). \label{eq29}
\end{equation}
Taking into account that when $r\rightarrow \infty $ the
expression in the brackets tends to unity we get the well known
result that $\bar \eta ={\dot M \over 3\pi}$ is a constant in the
nonrelativistic limit.

For our purposes we need an expression for kinematic viscosity
$\nu =\eta/\rho$. In this paper we employ the simplest possible
approach assuming that $\nu$ does not depend on the vertical
coordinate $\xi$ and use the standard \citet{S72} and \citet{SS73}
prescription for an estimate of this quantity
assuming that it is proportional to the product of a
characteristic disc scale height and sound speed, $c_s$, in the
disc
\begin{equation}
\nu \sim \alpha c_s h_{proper},  \label{eq30}
\end{equation}
where $h_{proper}$ is the proper scale height of the disc related
to a coordinate, $h$, of the disc upper boundary in our coordinate
system as $h_{proper}=K_2h$, and  $\alpha$ is the usual
Shakura-Sunyaev parameter. It is assumed to be a constant. From
the hydrostatic balance equation (\ref{eq25}) it follows that
$c_{s} \sim \sqrt{P/\rho} \sim U^{\varphi} h/r$. Substituting this
relation to (\ref{eq30}) we finally define $\alpha $ in such a way
that
\begin{equation}
\nu = \alpha K_2 U^{\varphi} h^{2}/r.  \label{eq30n}
\end{equation}
Using equations (\ref{eq29}) and (\ref{eq30n}) we obtain an
important relation
\begin{equation}
\Sigma h^{2} ={\dot M \over 3\pi \alpha}\left( {r^{1/2}\over
U^{\tau}(U^{\varphi})^{2}}{D\over K_1^3K_2^{5/2}} \right ). \label{eq30nn}
\end{equation}

Finally, we are going to use the ratio $\delta (r) =h/r$ in our
calculations. It can be easily obtained from the results of NT
provided that one takes into account a correction to the vertical
balance equation of NT found in RH. When the disc is gas pressure
dominated and the main source of opacity is determined by the
Thomson scattering, we have
\begin{equation}
\delta (r)= \delta_{*}
K_1^{3/5}K_2^{1/20}(U^{\tau})^{-4/5}D^{1/5}r^{1/20}, \label{eq31}
\end{equation}
and for the gas pressure dominated disc with the free-free
processes giving the main source of opacity we get
\begin{equation}
\delta (r)= \delta_{*}
K_1^{13/20}K_2^{1/8}(U^{\tau})^{-17/20}D^{3/20}r^{1/8},
\label{eq32}
\end{equation}
where $\delta_* \ll 1$ is a constant. Note that in both cases
$\delta=0$ at the marginally stable orbit.
It is also important to note that the disc opening angle $\delta(r)$
so defined does not depend on whether we use the Boyer-Lindquist
or the isotropic coordinates. Indeed, taking into account that
$h_{proper} = K_2 h$ and $R=K_2 r$ we have $\delta(r)=h/r = h_{proper}/R$.

The expressions for the flat NT model listed above are strictly
valid only when the radial drift velocity is small: $|U^{r}| \ll
c_{s}$. This inequality is, however, broken when $x=R-R_{ms}$ is
small, and, when $R$ is very close to the position of the last
stable orbit $|U^{r}|
> c_{s}$. Assuming that the NT expressions can also give order of
magnitude estimates even when $|U^{r}| \sim c_{s}$ let us estimate
a value of $x$, $x_{s}$, where we have $|U^{r}|=c_{s}$.

The expression for $|U^{r}|$ follows from equations (\ref{eq26})
and (\ref{eq30nn}),
\begin{equation}
|U^{r}|={3\alpha \over 2}{\delta^{2}\over D} K_1^2
U^{\tau}(U^{\varphi})^{2}\sqrt{R}, \label{eqmm1}
\end{equation}
while the speed of sound may be estimated as $c_{s}\sim
U^{\varphi}\delta$, see above. Assuming that the new radial
variable $x$ is small we have from (\ref{eq27})
\begin{equation}
D\approx {x^{2}\over 72}, \label{eqmm2}
\end{equation}
we represent the quantity $\delta $ as $\delta =
\delta_{ms}x^{2\epsilon}$, where $\delta_{ms}\sim \delta_{*}$, and
$\epsilon=1/5$, $3/20$ for the cases of the disc opacity dominated
by the Thomson scattering and free-free processes, respectively,
see (\ref{eq31}) and (\ref{eq32}). In the same limit we can set
the values of all variables entering (\ref{eqmm1}) and the
expression for $c_{s}$ excepting $\delta $ and $D$ equal to their
values at $R_{ms}$. We get
\begin{equation}
|U^{r}|/c_{s}\simeq A_{*}\alpha \delta_{*} x^{2(1-\epsilon)},
\quad A_{*}={1\over 48}K_{1}^{2}U^{\tau}U^{\varphi}\sqrt{R} \simeq
2.3 \cdot 10^{-2}, \label{eqmm3}
\end{equation}
where we use (\ref{eqmm2}). From this equation we obtain
\begin{equation}
x_{s} \simeq A_{*}^{1/2(1-\epsilon)} (\alpha
\delta_{*})^{1/2(1-\epsilon)}.\label{eqmm4}
\end{equation}
This equation tells that for typical values of $\alpha $ and
$\delta_{ms} \sim 10^{-2}-10^{-3}$ (see e.g. \citet{IIN98}, their
equation (1)~) the value of $x_{s}$ is quite small.

\section{Derivation of equations governing dynamics and
stationary shape of a twisted disc}\label{twist}

\subsection{Basic facts and simplifications}
As described above we use equations (\ref{eq19}) written
in the basis (\ref{eq13}-\ref{eq16}) to obtain equations
determining dynamics and stationary configurations of a twisted
disc and we take into account that our assumption about smallness
of the inclination angle $\beta $ allows us to decouple the set of
equation (\ref{eq19}) onto two subsets: 1) the subset of equations
describing the standard flat relativistic disc reviewed above (the
"standard subset") and 2) an additional set of equations
determining propagation of warped and twisted disturbances through
the disc and shapes of non-planar stationary configurations (we
call it later as a subset describing the disc twist). All
equations of this additional subset depend harmonically on the
angle $\psi $ (or $\varphi$) and can be obtained from the general
set (\ref{eq19}) by a procedure employing symmetry of different
terms in (\ref{eq19}) with respect to the vertical coordinate
$\xi$ since in the approximation we use all terms in this subset
are either even or odd with respect to the change $\xi \rightarrow
-\xi$. Namely, let us call the equations of (\ref{eq19}) having
the index $i=\tau, r, \varphi $ as "the horizontal part of
equations" and the equation with $i=\xi$ as "the vertical
equation". Then, all terms of the horizontal part having the odd
symmetry and all terms of the vertical equation having the even
symmetry belong to the subset of equations describing the disc
twist while all terms with the opposite symmetries belong to the
standard subset.

There are additional simplifications determined by the fact that
we consider only the thin discs $h/r \ll 1$ and slowly rotating
black holes with $a \ll 1$.

The first assumption allows us to consider only terms of order of
$h/r$ in the horizontal part and terms of zero and second order in
$h/r$ in the vertical equation. In fact, as we are going to show
there is only one zero order term in the vertical equation
describing precession of the disc rings due to the Lense-Thirring
effect (the so-called Lense-Thirring or gravitomagnetic
precession). It is, therefore, proportional to $a$. Thus, this
zero order term may also be classified as a "small" one due to our
second assumption. A further simplification follows from the fact
that the pressure $p$, the thermal density $\epsilon_{th}$ and the
dynamic viscosity  $\eta$ are small in comparison with $\rho$: $p,
\epsilon_{th}, \eta \sim (h/r)^{2}\rho$. Thus, these quantities
enter in the horizontal part only in expressions containing
differentiation over $\xi$ in order to provide terms linear in
$h/r$. In the vertical equation we set $\epsilon = \rho $ due to
the same assumption. This results in absence of $\epsilon_{th}$ in
all our expressions used below, see Appendix B.

The second assumption allows us to suppose that a characteristic
evolution time scale of propagation of twisted and warped
disturbances through the disc, $t_{tw}$, is much larger than the
disc dynamical time scale $t_{d}\sim \Omega^{-1}$. Indeed, for a
non-rotating black hole with $a=0$ the disc rings inclined with
respect to each other can change their mutual orientations only
due to interactions determined by either pressure or viscous
forces\footnote{Let us remind that we neglect self-gravity of the
disc.}. Due to our first assumption these interactions are small
and the corresponding time scales must be proportional to an
inverse power of $h/r$. For a slowly rotating black hole there is
an additional characteristic time scale, $t_{LT}$, determined by
the gravitomagnetic precession of the rings. It may be estimated
as $t_{LT}\sim a^{-1} r^{3} \gg t_{d}$, see e.g. equation
(\ref{2A11}). Due to the small ratios $t_{d}/t_{tw}$
and $t_{d}/t_{LT}$ the time derivatives of the Euler angles, and,
accordingly, the variable $U$ defined in (\ref{eqn8}) can be
neglected in all expressions entering in the horizontal part.

As was first noted by \citet{PP83}, hereafter PP, for the
Newtonian problem in order to get a self-consistent set of
equations describing the disc twist one must take into account
perturbations of velocity, gas density and pressure determined by
the twisted disturbances in the disc. In our relativistic
generalisation we must use perturbations of four-velocity, and,
accordingly, assume that
\begin{equation}
U^{i}=U^{i}_{0}+v^{i}, \quad \rho=\rho_{0}+\rho_{1}, \quad p = p_0 + p_1 \label{eq33}
\end{equation}
where the quantities with index $(0)$ are determined by the
equations of the flat disc model described above. Note that in a
self-consistent approach we should consider perturbations of the
dynamical viscosity, $\eta=\eta_0+\eta_1$, and radiation flux,
$q^i = q^i_0 + q^i_1$, as well. However, it can be shown that
contribution of terms proportional to the unperturbed part of
radiation flux, its perturbation and perturbation of the dynamical
viscosity, is negligible in our equations at the required order of
accuracy.
This is related to the fact that the quantities $\eta_1$ and
$\frac{\partial q^\xi}{\partial \xi}$ are proportional to
$(h/r)^2$ and $q^r,q^\varphi\propto (h/r)q^\xi$. As we discuss
above the terms proportional to $(h/r)^2$ give a contribution
to the horizontal part of our equations being differentiated
over $\xi$ and in the vertical part being multiplied by
terms of zero order in $h/r$. A direct examination of
our set of the perturbed equations shows that this, in fact,
does not happen in the linear approximation in angle $\beta$.
Also note that the pressure perturbation $p_1$ does not
enter in our set of final equations describing a twisted disc, see the next Section.

Since both $U^{i}$ and
$U^{i}_{0}$ obey the same normalisation condition
$U_{i}U^{i}=U_{0i}U^{i}_{0}=1$ the velocity perturbations $v^{i}$
are orthogonal to $U_{0}^{i}$: $v^{i}U^{i}_{0}=0$, giving
$v^{0}=(U^{\varphi}_{0}/U^{0}_{0})v^{\varphi}$. The "horizontal"
velocity perturbations $v^{0}, v^{\varphi}, v^{r}$ are odd
functions of $\xi$ while the "vertical" perturbation $v^{\xi}$ is
an even function of the same coordinate.

The horizontal part of the velocity perturbation is directly
determined from our equations for the disc twist, see below.
Determination of the vertical part, $v^{\xi}$ requires some
additional consideration. Namely, as was first shown for the
Newtonian problem by \citet{Hat81}, $v^{\xi}$ consists of two
parts: $v^{\xi}=d\xi/dt+v^{\xi}_b$, the first one is determined by
a change of the coordinate $\xi$ with time of a given gas element
while the second one is due to the non-coordinate character of our
basis vectors (\ref{eq13}-\ref{eq16}). This results in evolution
of projection of the velocity vector $v^{i}$ onto the vector ${\bf
e}_{\xi}$ even when a gas element has a fixed value of $\xi$
provided that either the Euler angles are changing with time or
there is a non-zero drift velocity component $U^{r}_0$. In order
to find $v^{\xi}_b$ let us assume that $d\xi/dt=0$. In this case
we have $v^{\xi}=\mbox{\boldmath$\omega$}^{\xi}/ds$, 
where $\mbox{\boldmath$\omega$}^{\xi}$ is given by equation
(\ref{eq12}), and $ds$ is the
line element. Using this equation, taking into account that
$U_0^{\tau}=K_1d\tau/ds$ and $U_0^{r}= K_2dr/ds$, see equations
(\ref{eq9}) and (\ref{eq10}), respectively, we get
\begin{equation}
 v^\xi = r U_0^\tau \frac{K_2}{K_1} U + r U_0^r W.
 \label{eq34}
\end{equation}
In what follows we set $d\xi/dt=0$. This requirement is related to
a specific freedom of choice between different twisted coordinate
systems corresponding to the same physical situation\footnote{This
is analogous to the freedom of choice of a gauge in Quantum Field
Theory and General Relativity.}. Indeed, in order to construct our
twisted coordinate system we add two Euler angles to the set of
the state variables describing our physical system while the
number of dynamical equations remains the same. Therefore, in the
twisted coordinate system there is a freedom of choice between
different sets of dynamical variables satisfying the same
equations, which must be fixed by an additional requirement. As
was discussed in \citet{IP08} the requirement $d\xi/dt=0$ fixes
this freedom and leads to a choice of the most appropriate twisted
coordinate system, where perturbations of the state variables are
minimal\footnote{Note that we assume in this paper that there is
no surface forces acting on the disc, for a more general case this
requirement does not hold, see \citet{IP08} for details.}.

Since the unperturbed parts of our variables and the perturbations
enter in a different way in all equations below, we omit for
simplicity the index $(0)$ characterising  the unperturbed
variables from now on.

The horizontal part of our equations and the vertical equation are
shown in Appendix B, see equations (\ref{2A1}-\ref{2A3}) and
equation (\ref{2A4}), respectively. They have a rather complicated
structure. However, they can be further simplified. Indeed, a
simple analysis shows that  all terms in these equations
proportional to the rotational parameter $a$ excepting the
gravitomagnetic term - the second term in square brackets on the
right hand side of (\ref{2A4}) can be neglected.  These terms lead
to terms in our final equation (\ref{eq38}) proportional to a
product of two small parameters - $\propto a(h/r)^{2}$. Thus, we
can consider an effectively Schwarzschild problem with inclusion
of only one term determined by rotation of the black hole.
Additionally, the time derivatives of $\rho_{1}$ and $v^{\xi}$ can
also be neglected since, similar to the Newtonian problem, they
result in terms proportional to $(t_{d}/t_{tw})(h/r)^{2}$ in the
same equation. On the other hand, the time derivatives of the
perturbed "horizontal" velocities $v^{r}$ and $v^{\varphi}$ should
be retained in a certain combination, see e.g. \citet{PL95}, DI
and below. The reason for this is the well known degeneracy of the
Keplerian gravitational potential resulting in closed orbits of
particles around a Newtonian source of gravity. Owing to this
degeneracy a certain combination of terms proportional to the time
derivatives (we call it later the "resonance combination") plays
an important role at large distances from the black hole, where
the gravitational potential is approximately the Newtonian one
provided that $\alpha \ll 1$.

\subsection{The subset of equations describing the disc twist}

Taking into account the simplifications discussed above let us
consider equations of Appendix B. At first let us deal with the
'horizontal' part of the problem. It turns out that it is more
instructive to use certain linear combinations of equations
(\ref{2A1}) and (\ref{2A3}) instead of themselves. Neglecting the
terms proportional to $a$ and $\dot \rho_1$ in these equations and
taking the linear combination $(\ref{2A1})\times \left (
\frac{(U^\tau)^2 + (U^\varphi)^2}{U^\tau}\right ) -
(\ref{2A3})\times \left ( 2U^\varphi \right ) $ we remove the time
derivative of $v^{\varphi}$ from the result, thus obtaining
\begin{equation}
U^\varphi
\partial_\varphi \rho_1 + \frac{1}{(U^\tau)^2} \rho
\partial_\varphi v^\varphi + \frac{U^\tau}{K_2^2}
\frac{\partial}{\partial r} \left ( r K_2^2 \frac{\rho
v^r}{U^\tau} \right )=  \xi  U^\varphi \rho
\partial_\varphi W + \frac{U^\varphi}{(U^\tau)^2} (\partial_\xi
T_{\nu}^{\varphi \xi}-rW
\partial_\xi T_{\nu}^{r\varphi }).
\label{eq35}
\end{equation}
In the Newtonian limit, when $r\rightarrow \infty $, equation
(\ref{eq35}) reduces to the continuity equation. Another equation
reducing in the Newtonian limit to the $\varphi $ component of
perturbed Navier-Stokes equations can be obtained from (\ref{2A1})
and (\ref{2A3}) taking another linear combination
$((\ref{2A3})\times U^\tau - (\ref{2A1})\times U^\varphi)/\rho$
with the result
\begin{equation}
\frac{K_2}{K_1} \dot v^\varphi +
\frac{1}{r}\frac{U^\varphi}{U^\tau} \partial_\varphi v^\varphi +
\left ( \frac{\partial_r U^\varphi}{U^\tau} +
\frac{K_1^\prime}{K_1} \frac{U^\tau}{U^\varphi} \right ) v^r +
\frac{1}{\rho U^\tau}(\partial_\xi T_{\nu}^{\varphi \xi}-rW
\partial_\xi T_{\nu}^{r\varphi }) = 0. \label{eq36}
\end{equation}
With our simplification being adopted equation (\ref{2A2}) has the
form
\begin{equation}
 \frac{K_2}{K_1} U^\tau \dot v^r +
\frac{U^\varphi}{r} \partial_\varphi v^r - 2\frac{K_1^\prime}{K_1
U^\varphi} v^\varphi +{1\over \rho}\partial_\xi T_{\nu}^{r\xi} = W r
\frac{\partial_\xi p}{\rho}. \label{eq37}
\end{equation}
It is easy to show that it reduces to the $r$ component of
perturbed Navier-Stokes equations when $r\rightarrow \infty $.

The vertical equation (\ref{2A4}) plays a special role in our
analysis. In fact, as we see later it most important for
determination of the dynamics and form of twisted configurations
of accretion discs. In the approximation we use, i.e. setting all
terms proportional to $a$ excepting the gravitomagnetic term in
(\ref{2A4}), and, accordingly, in (\ref{2A5}-\ref{2A7})  to zero
we can bring this equation to the so-called divergent form, which
reflects symmetries of the Schwarzschild space-time with respect
to spacial rotations. This is analogous to the Newtonian problem,
see PP and DI. For that we would like to express the density
perturbation $\rho_1$ and the velocity perturbation $v^{\varphi}$
entering (\ref{2A4}) through $v^{r}$ using equations
(\ref{eq35}-\ref{eq37}). Since the resonance combination of terms
containing the time derivatives does not enter in the vertical
equations we can set $\dot v^{\varphi}=\dot v^{r}=0$ when
operating with equations (\ref{eq35}-\ref{eq37}). After this being
done we express the term containing $\partial_{\varphi} W$ in
(\ref{eq35}) through the velocity perturbations using equations
(\ref{eq25}) and (\ref{eq37}), and substitute the result in
(\ref{eq35}). We express the velocity perturbation $v^{\varphi}$
entering explicitly in (\ref{eq35}) and (\ref{2A4}), in terms of
$v^{r}$ with help of (\ref{eq36}), and substitute it in these
equations. Then, we use the relation obtained from (\ref{eq35}) to
express the density perturbation through $v^r$ and the viscous
terms and substitute the result in (\ref{2A4}). The resulting
expression is integrated over $\xi$, where we note that the term
containing the derivative of the perturbed pressure, $p_1$,
vanishes after integration. This expression also contains
derivatives over $\xi$ in the viscous terms (\ref{2A5}-\ref{2A6})
under the integrals. These are integrated by parts taking into
account that the corresponding surface terms are equal to zero.
After a rather tedious but straightforward calculation along the
lines outlined above we obtain from (\ref{2A4}) a rather simple
equation
\begin{multline}
\Sigma U^\tau U^\varphi \left \{\partial_\varphi U - a\frac{K_1
K_3}{K_2^2} Z \right \} + \partial_\varphi W \frac{K_1}{K_2} \left
\{\Sigma U^\varphi U^r + \bar T_{\nu}^{r\varphi} \right \}=\\
-\frac{1}{ 2r^2  K_2^4} \int d\xi\, \{\, \partial_r (\xi r K_1
K_2^3 U^\varphi \rho \partial_\varphi v^r + r^2 K_1 K_2^3 T^{r\xi}
), \label{eq38}
\end{multline}
where  we remind that $\Sigma= \int \rho\, d\xi $ is the surface
density and the bar stands for the quantities integrated over
$\xi$.

In Appendix C we show that when $a=0$ equation (\ref{eq38})
may be brought in a special divergent form determined by rotational
symmetries of the Schwarzschild space-time.

Provided a background model of a flat relativistic disc is
specified equations (\ref{eq36}-\ref{eq38}) form a complete set.

\subsection{A model case of an accretion disc having isothermal
density distribution in the vertical direction}

In order to transform the set of equations (\ref{eq36}-\ref{eq38})
to a simpler form we should specify a dependence of density $\rho
$ on $\xi$. Since an accurate treatment of the vertical
distribution of the density is not important for our purposes we
would like to consider the simplest possible case of an isothermal
disc
\begin{equation}
\rho = \rho_c \,{\rm exp} (-\frac{\xi^2}{2h^2}), \label{eq39}
\end{equation}
where the central density $\rho_{c}$ and the disc half-height $h$
are functions of $r$.

Similar to the Newtonian problem it is easy to see that
distributions of the velocity perturbations of the form
\begin{equation}
v^\varphi = \xi (A_1 \sin \varphi + A_2 \cos \varphi)\quad v^r =
\xi (B_1 \sin \varphi + B_2 \cos \varphi)\label{eq40}
\end{equation}
satisfy equations (\ref{eq36})  and (\ref{eq37}) provided that the
kinematic viscosity $\nu$ does not alter with height and the
amplitudes $A_1$, $A_2$, $B_1$ and $B_2$ are assumed to be
functions of the time and radial coordinates.

Introducing the complex notation
\begin{equation}
{\bf A} = A_2 + i A_1, \quad {\bf B} = B_2 + i B_1 \quad
\mbox{and} \quad {\bf W} = \Psi_1 + i \Psi_2 = \beta e^{i\gamma}
\label{eq41}
\end{equation}
and using equations (\ref{eq25}), (\ref{eq30n}), (\ref{eq40}),
(\ref{2A5}) and (\ref{2A7}) we can rewrite (\ref{eq36}) and
(\ref{eq37}) in the form
\begin{equation}
\dot {\bf A} - (i-\alpha)\Omega {\bf A} +
\frac{\kappa^2}{2\tilde\Omega} {\bf B} = -\frac{3}{2} i \alpha K_1
(U^\tau)^2 U^\varphi \Omega {\bf W}^\prime \label{eq42},
\end{equation}
\begin{equation}
\dot {\bf B} -  (i-\alpha)\Omega {\bf B} - 2\tilde \Omega {\bf A}
= - (i+\alpha) U^\varphi\Omega\, {\bf W}^\prime \label{eq43},
\end{equation}
respectively, where
\begin{equation}
\kappa^2 = R^{-3} \left (1-\frac{6}{R} \right ), \quad
\tilde\Omega = \frac{R - 3}{R^2 (R-2)^{1/2}}.\label{eq44}
\end{equation}
Note that $\kappa $ is the relativistic epicyclic frequency for a
slightly perturbed circular orbit in the Schwarzschild space-time,
see e.g. \citet{AG81,Kat90}.

As we have mentioned above equations (\ref{eq42}), (\ref{eq43})
contain a degeneracy in the Newtonian limit $r\rightarrow \infty
$, for a formally inviscid disc, due to the well known Keplerian
resonance between the mean and epicyclic motions. It is due to
this resonance the "small" terms $\dot {\bf A}$ and $\dot {\bf B}$
should be retained in these equations. In order to clarify the
physical meaning of this resonance let us consider the limit
$r\rightarrow \infty $ and set $\alpha=0$ in equations
(\ref{eq42}) and (\ref{eq43}). We get
\begin{equation}
\dot {\bf A} - i\Omega {\bf A} + {\Omega \over 2} {\bf B} =0
\label{eq45}
\end{equation}
from equation (\ref{eq42}), and
\begin{equation}
\dot {\bf B} -  i\Omega {\bf B} - 2\Omega {\bf A} = - i
U^\varphi\Omega\, {\bf W}^\prime \label{eq46},
\end{equation}
where $U^{\varphi}$ is assumed to have its Keplerian value,
$U^{\varphi}=r^{-1/2}$. Multiplying (\ref{eq45}) by $2i$ and
adding the result to (\ref{eq46}), we see that the terms
proportional to $\Omega $ on the left hand sides of (\ref{eq45})
and (\ref{eq46}) cancel each other, and we obtain
\begin{equation}
\dot {\bf B} +2i\dot {\bf A} = -i U^\varphi\Omega\, {\bf W}^\prime
\label{eq47}
\end{equation}
On the other hand we must have $(|\dot {\bf B}|, |\dot {\bf A}|)
\sim (t_{d}/t_{tw}) (|\Omega  {\bf A}|,|\Omega  {\bf B}|)$, where
we remind that $t_d\sim \Omega^{-1}$ is the characteristic
dynamical time scale, $t_{tw}$ is the characteristic time scale of
evolution of twisted disturbances and $t_{d}/t_{tw} \ll 1$. Taking
this fact into account we see that the term on the right hand side
of (\ref{eq47}) is of order of $(t_{d}/t_{tw}) |\Omega {\bf A}|$.
Therefore, it can be neglected in the leading approximation in
(\ref{eq46}), and we get either from this equation or from
(\ref{eq45}) an algebraic relation between ${\bf A}$ and ${\bf
B}$:
\begin{equation}
{\bf B}= - 2i {\bf A}, \label{eq48}
\end{equation}
which is valid only in the leading order in $t_{d}/t_{tw}$. Thus,
in this order the perturbed velocities can be described by only
one complex amplitude, either ${\bf A}$ or ${\bf B}$\footnote{As
discussed in e.g. DI the relation (\ref{eq48}) follows from the
fact that slightly perturbed circular orbits of free particles in
the Newtonian potential are ellipses with a small eccentricity.}.
Let us show that the same property approximately holds for the
general relativistic problem as well. For that let us consider a
linear combination of (\ref{eq42}) and (\ref{eq43}) with
coefficients chosen in such a way that the term proportional to
$\bf{A}$ is eliminated
\begin{equation}
\dot {\bf B} - \frac{2\tilde\Omega}{(i-\alpha)\Omega} \dot {\bf A}
= \left [ 1 + \frac{\kappa^2}{(i-\alpha)^2 \Omega^2} \right ]
(i-\alpha)\Omega {\bf B} - \left [ (i+\alpha)U^\varphi \Omega -
\frac{3i\alpha}{i-\alpha} K_1 (U^\tau)^2U^\varphi \tilde \Omega
\right ] {\bf W}^\prime . \label{eq49}
\end{equation}
The combination on the left hand side plays a role in dynamics of
the disc only when we consider a non-relativistic, low viscosity
disc with $\alpha < h/r$ while in the opposite limits it can be
set to zero. Therefore, we can consider all background quantities
in this combination equal to their Newtonian limits, set
$\alpha=0$ there and use equation (\ref{eq48}) to express ${\bf
A}$ through ${\bf B}$ on the right hand side. We get
\begin{equation}
\dot{\bf B} =  { 1\over 2} \left \lbrace \left [ 1 +
\frac{\kappa^2}{(i-\alpha)^2 \Omega^2} \right ] (i-\alpha)\Omega
{\bf B} - \left [ (i+\alpha)U^\varphi \Omega -
\frac{3i\alpha}{i-\alpha} K_1 (U^\tau)^2U^\varphi \tilde \Omega
\right ] {\bf W}^\prime \right \rbrace \label{eq50}
\end{equation}
Equation (\ref{eq50}) describes the horizontal part of the problem
in the leading order in $t_{d}/t_{tw}$, for all possible values of
$\alpha $ and $r$. It is one of two final equations describing a
fully relativistic twisted disc around a slowly rotating black
hole.

In order to get the second final equation describing our problem
let us consider the vertical equation (\ref{eq38}). At first we
use equations (\ref{eq39}-\ref{eq41}) and (\ref{2A7}) to rewrite
it in terms of the complex variables and perform integration over
$\xi$. After this being done we get expression proportional to
$\bar \eta $ and $\Sigma h^{2}$ under the derivative over $r$. We
substitute (\ref{eq29}) and (\ref{eq30n}) and take the constant
factors out of the derivative. Then, we use equation (\ref{eq26})
to express $U^{r}$ in terms of $\dot M$ and the explicit form of
$\bar T^{r\varphi}_{\nu}$ given by (\ref{eq26nn}). After the
resulting expression is divided over $\Sigma $ one can see that it
contains $\dot M$ and $\Sigma $ only in the combination $\dot
M/\Sigma $. This combination is expressed in terms of $\delta=h/r$
with help of (\ref{eq30nn}). We finally obtain
\begin{multline}
\dot {\bf W} - i\Omega_{LT} {\bf W} + \frac{3}{2} \alpha \delta^2
\frac{K_1^2}{K_2} U^\varphi \left ( U^\tau - K_1 (r K_2)^{1/2}
\frac{U^\varphi}{D} \right ) {\bf W}^\prime = \\
\frac{\delta^2 K_1^3 U^\varphi}{2 r^{1/2} K_2^{3/2} D} \frac
{\partial}{\partial r} \left \{ r^{3/2} K_2^{1/2} \frac{D}{K_1^2
U^\tau U^\varphi} (\,\,(i+\alpha){\bf B} +\alpha U^\varphi {\bf
W}^\prime\,) \right \}. \label{eq51}
\end{multline}

Equations (\ref{eq50}) and (\ref{eq51}) form a complete set.

\section{A stationary twisted disc}

Non-stationary solutions of equations (\ref{eq50}) and
(\ref{eq51}) will be discussed in a separate publication. In this
paper we would like to consider a stationary problem setting time
derivatives in (\ref{eq50}) and (\ref{eq51}) to zero. Then, we
express ${\bf B}$ through ${\bf W}^{\prime}$ using (\ref{eq50})
and substitute the result in (\ref{eq51}) to obtain
\begin{equation}
\frac{K_1}{R^{1/2} D} \frac{d}{d R} \left ( \frac{R^{3/2} D}{K_1
U^\tau} f^{*}(\alpha, R) \frac{d {\bf W}}{d R} \right ) - 3\alpha
U^\tau (1-D^{-1}) \frac{d {\bf W}}{d R} + \frac{4ia}{\delta^2
K_1^3 R^3 U^\varphi} {\bf W} = 0, \label{eq52}
\end{equation}
where $*$ stands for the complex conjugate,
\begin{equation}
f(\alpha, R) = (1 + \alpha^2 - 3i\alpha K_1^2)\,\,
\frac{R(i-\alpha)} {\alpha R(\alpha+2i)-6} + \alpha, \label{eq53}
\end{equation}
we use the coordinate $R$ as an independent variable, and the
explicit expression for the gravitomagnetic frequency
$\Omega_{LT}$ (\ref{2A11}) is used. It is easy to see that in the
non-relativistic limit $R\rightarrow \infty $ and $K_1\rightarrow
1$ the function $f(\alpha, R)$ tends to the function $f(\alpha
)={(2+6i\alpha +i/\alpha)\over (\alpha + 2i)}$ introduced in paper
\citet{KP85}, hereafter KP.

A character of solutions to equation depends on only two
independent parameters - $\alpha $ and
\begin{equation}
\tilde \delta =\delta_{*}/\sqrt {|a|} \label{eq57}
\end{equation}

\subsection{Limiting cases}

A number of authors considered equations describing a shape of a
stationary twisted disc far from a rotating black hole. They used
either a purely Newtonian analysis  with the gravitomagnetic term
treated as originating from a Newtonian force responsible for
precession of the disc rings (e.g. PP, KP) or took into account
certain post Newtonian terms, which may play a significant role in
determining  of the stationary disc configurations (e.g. \citet{II97}, hereafter II;
\citet{LOP02}).
Obviously, equation (\ref{eq52})
should be reduced to equations exploited by previous authors in
the respective limits.

In particular, in the paper II a low viscosity $(\alpha \ll 1)$
stationary twisted disc was considered and a simple equation
describing the stationary configurations was derived taking into
account only the most important post-Newtonian correction to the
first term on the right hand side of (\ref{eq52}). In the
approximation used by II we neglect the second term on the right
hand  side of (\ref{eq52}), which is proportional to $\alpha $ and
set the values of all quantities there equal to their Newtonian
limits (i.e. $U^{\tau}=1$, $U^{\varphi}=R^{-1/2}$, $D=1$ and
$K_1=1$) and assume that $\alpha=0$ everywhere excepting the
so-called 'resonance' denominator in the first term in
(\ref{eq53}), which should contain a term proportional to $\alpha$
and a first relativistic correction. In this approximation we have
$f(\alpha, R)={1\over 2\alpha }{1\over (1 +{3i\over \alpha R})}$
and obtain from equation (\ref{eq52})
\begin{equation}
{d\over dx_1}{1\over (1 -3i \alpha^{-1} x_1^2)} {d\over dx_1}{\bf
W} +{32 i\alpha a x_1\over \delta^{2}}{\bf W}=0,\label{eq54}
\end{equation}
where $x_1=R^{-1/2}$. This equation coincides with equation (33)
of II.

PP derived dynamical equations describing a Newtonian twisted
accretion disc in the limit of relatively large viscosity $\alpha
\gg \delta $. Stationary solutions to these equations were
examined in the paper KP, for a disc with isothermal distribution
of density with height. In order to obtain a fully Newtonian limit
of (\ref{eq52}) and compare it to equation used by KP  we set
$U^{\tau}=1$, $U^{\varphi}=R^{-1/2}$, and $K_1=1$ in (\ref{eq52})
and (\ref{eq53}). We also introduce the independent variable
$x_2=\sqrt{R_{ms}}x_1=\sqrt{R_{ms}/R}$, use the form of $D=1-x_2$
employed by KP and parametrisation of $\delta $,
$\delta=(H_{*}/R_{ms})x_2^{-2(g-1)}\sqrt D$, where $g$ is a
parameter as in KP. In this way we obtain from (\ref{eq52})
equation studied by KP
\begin{equation}
x_2^{(9-4g)}\left (f^*(\alpha){d\over dx_2}\left [(1-x_2){d\over dx_2}{\bf
W}\right ]-6\alpha {d\over dx_2}{\bf W} \right )+16iax_2^6{\sqrt R_{ms} \over
H_{*}^2}{\bf W}=0.\label{eq55}
\end{equation}

\subsection{An almost inviscid twisted disc}

An important limiting case of (\ref{eq52}) is obtained by formally
setting $\alpha=0$ in this equation. When the disc gas rotates in
the same sense as the black hole and, accordingly, $a
> 0$ the behaviour of the disc inclination angle differs
drastically from the standard picture of the well known
Bardeen-Petterson effect, where it is expected that the disc
aligns with the black hole equatorial plane at sufficiently small
radii. Contrary to this picture when considering the shape of low
viscosity twisted disc at large distances $R \gg R_{ms}$ II found
a phenomenon of spatial oscillations of the disc inclination angle
$\beta $ for the case of $a > 0$ while in the opposite case the
disc does align with the black hole equatorial plane. Our fully
relativistic analysis allows us to consider this effect in more
details and show that the amplitude of the oscillations can be
significantly amplified close to the last stable orbit.

Setting $\alpha=0$ in (\ref{eq52}) we get
\begin{equation}
{d\over dR} \left (b {d\over dR} {\bf W}\right )+ \lambda {\bf W} =0,
\label{eqn1}
\end{equation}
where
\begin{equation}
b= {R^{5/2}D\over K_1 U^{\tau}}, \quad \lambda = {24 a D\over
\delta^2 K_1^4 U^{\varphi} R^{5/2}}. \label{eqn2}
\end{equation}
Note that equation (\ref{eqn1}) depends parametrically only on
$\tilde \delta $. Since this equation has only real coefficients
its solution may be taken to be real. Thus, in this limit only the
angle $\beta $ changes with radius while $\gamma $ stays
unchanged.

It is also important to note that the influence of the black hole
gravitomagnetic force on the disc shape is most significant when
$\tilde \delta \ll 1$. Also, it is most natural to suppose that
the rotation parameter $a > 0$. The case of $\tilde \delta \ll 1$
and $a > 0$ is considered in more details below. Note, however,
that the bulk of our expressions below are formally valid for the
case $a < 0$ as well. Although in the case considered below we may
choose ${\bf W}=\beta $ we continue to use ${\bf W}$ to keep our
notation uniform.

\subsubsection{The shape of an almost inviscid stationary twisted
disc close to $R_{ms}$}

At first let us discuss the form of solutions to (\ref{eqn1})
close to the last stable orbit assuming that $x=R-R_{ms} \ll 1$.
For that we change the independent variable to $x$ in (\ref{eqn1})
and assume that all quantities entering this equation, which have
finite values at $R_{ms}$ are equal to these values. Thus, we take
into account only the dependencies of $D$ and $\delta $ on $x$ in
such approximation. We use equation (\ref{eqmm2}), which tells
that $D\approx x^2/72$ close to $R_{ms}$.

We parametrise the dependency of $\delta $ on $x$ as
$\delta=\delta_{ms}x^{2\epsilon}$, with $\delta_{ms}$ being
proportional to $\delta_{*}$. We are going to consider only the
values of $\epsilon < 1/2$, which physically corresponds to the
condition that the surface density $\Sigma $ tends formally to
zero when $R\rightarrow R_{ms}$. Equations (\ref{eq31}) and
(\ref{eq32}) tell that $\epsilon=1/5$,
$\delta_{ms}=0.37\delta_{*}$ and $\epsilon=3/20$,
$\delta_{ms}=0.51\delta_{*}$, for the opacity dominated by the
Thomson scattering and the free-free processes, respectively.

In this way we get from (\ref{eqn1})
\begin{equation}
{1\over x^2}{d\over dx}x^{2}{d\over dx}{\bf W}+ \chi
x^{-4\epsilon}{\bf W}=0, \label{eq56}
\end{equation}
where
\begin{equation}
\chi ={24 a U^{\tau}\over R^5K_1^3U^{\varphi}\delta^{2}_{ms}}.
\label{eq56a}
\end{equation}
is evaluated at $R_{ms}$. 
Using the parameter $\tilde \delta $ given by (\ref{eq57}) we can
rewrite (\ref{eq56a}) in the form $\chi = \tilde \chi {\tilde
\delta}^{-2}$, where $\tilde \chi_{Th}=8.4\cdot 10^{-2}$ and
$\tilde \chi_{ff}=4.4\cdot 10^{-2}$, the indices $Th$ and $ff$
stand for the cases of the Thomson and free-free opacities,
respectively. When $a
> 0$ the solution to equation (\ref{eq56}) regular at $R=R_{ms}$ may
be expressed through the Bessel function, $J_{\nu}(z)$, as
\begin{equation}
{\bf W}=Cx^{-1/2}J_{1/2(1-2\epsilon)}(z), \label{eq58}
\end{equation}
where
\begin{equation}
z={\sqrt \chi \over (1-2\epsilon)}x^{1-2\epsilon}.\label{eq59}
\end{equation}
It is easy to see that the solution (\ref{eq58}) tends to a
non-zero constant when $z \rightarrow 0$. This fact may be used to
express the constant $C$ through the value of ${\bf W}$ at $x=0$,
${\bf W}_{0}$, as
\begin{equation}
C=\Gamma \left ({3-4\epsilon \over 2(1-2\epsilon)} \right ) \left ({\sqrt \chi \over
2(1-2\epsilon )}\right )^{-1/2(1-2\epsilon)}{\bf W}_{0}, \label{eq60}
\end{equation}
where $\Gamma (x)$ is the gamma function and we use the well known
approximate expression for $J_{\nu}(z)$ at small $z$. Since the
solution contains the Bessel functions of a real argument it
describes oscillations of the inclination angle $\beta
$\footnote{Let us stress again that in the case of $a < 0$ the
solution contains the Bessel function of an imaginary argument,
which describes a monotonic decrease of the inclination angle of
the disc with decrease of $x$.}.

Let us estimate a change of the amplitude of these oscillations
over the range $0 < x <1$, in the limit of $\tilde \delta \ll 1$.
As follows from (\ref{eq56a}) and (\ref{eq59}) in this limit the
variable $z $ takes values $\gg 1$ even for $x < 1$ and,
therefore, the region can be divided on two domains: $z < 1$ and
$z > 1$. In the domain $z < 1$ the value of ${\bf W}$
monotonically decreases with $x$, and, typically, it decreases in
six times before oscillations of the inclination angle begin.
Thus, we have ${\bf W}_{0}/{\bf W}(z\sim 1)\sim 6$. When $z\gg 1$
we use the asymptotic expansion of the Bessel function in powers
of $z^{-1}$ simplifying (\ref{eq58})  to
\begin{equation}
{\bf W}\approx  C\sqrt{2\over \pi xz}\cos \left (z -{\pi \over
2}{1-\epsilon \over 1- 2\epsilon }\right ). \label{eq61}
\end{equation}
Assuming that this expression is approximately valid even when
$x\sim 1$ or $z\sim 1$ and that $1-2\epsilon \sim O(1)$, we can
roughly estimate
\begin{equation}
{\bf W}(x \sim 1) \sim \chi^{-(1-\epsilon)/2(1-2\epsilon)}{\bf
W}(z\sim 1). \label{eq62}
\end{equation} Using the quantities $\tilde \delta $ and $\tilde \chi$ in
(\ref{eq62}) and remembering that ${\bf W}_{0}/{\bf W}(z\sim
1)\sim 6$ we obtain
\begin{equation}
{\bf W}_0/ {\bf W}(x\sim 1)  \sim \,\left( \,6{\tilde
\chi}^{(1-\epsilon)/2(1-2\epsilon)} \,\right )\, \tilde \delta
^{-(1-\epsilon)/(1-2\epsilon)}. \label{eq63}
\end{equation}
Since $6{\tilde \chi}^{(1-\epsilon)/2(1-2\epsilon)}\sim O(1)$ for
the parameters we use we set this factor to unity in our
expressions below.

From equation (\ref{eq63}) it follows that when $\alpha
\rightarrow 0$ and $\tilde \delta \ll 1$ a rise of the amplitude
of the oscillation with decrease of $x$ near the marginally stable
orbit can be quite large.

\subsubsection{The shape of the disc at large radii: the case of $a >
0$ and $\tilde \delta \ll 1$}

II showed that when the limit $\tilde \delta \ll 1$ is considered
a low viscosity disc inclined at large distances with respect to
the equatorial plane gets twisted at a characteristic scale
\begin{equation}
R_2\sim {\tilde \delta}^{-4/5} \gg R_{ms}.
\label{eq63a}\end{equation}
 At such
large scales we can use equation (\ref{eq54}) considering the
limit $\alpha \rightarrow 0$ there. In this limit we get
\begin{equation}
x_1{d^2\over dx_1^2}{\bf W}-2{d\over dx_1}{\bf W}+96{\tilde
\delta}^{-2}x_1^4{\bf W}=0, \label{eq64}
\end{equation}
where we remind that $x_1=R^{-1/2}$ and that we consider the case
$a > 0$. Note that following II we neglect a very slow dependence
of $\delta $ on $R$ for $R \gg 1$ and set $\delta=\delta_{*}$ in
(\ref{eq64}). The solution to (\ref{eq64}) can be expressed
through the Bessel functions
\begin{equation}
{\bf W}=x_1^{3/2}(A_1J_{-3/5}(z_1)+A_2J_{3/5}(z_1)), \label{eq65}
\end{equation}
where
\begin{equation}
z_1={8\over 5}\sqrt 6{\tilde \delta}^{-1}x_1^{5/2},\label{eq65a}
\end{equation}
and $A_1$ and $A_2$ are constants \footnote{Note that an analogous
equation of II contains a misprint. Namely, the power of their
variable $y_2$ (which is equal to our variable $z_1$) in front of
the square brackets in their equation (37) should be $3/5$ and not
$3/2$ as in the text.}. The limit of large $R \gg R_2$ corresponds
to the limit of small $z_1$. In this limit the first and second
terms in the brackets in  (\ref{eq65}) being multiplied by
$x_1^{3/2}$ tend to a non-zero constant and to zero, respectively.
This fact may be used to express the constant $A_1$ through the
asymptotic value of ${\bf W}$ at infinity, ${\bf W}_\infty$, as
\begin{equation}
{\bf W}_\infty ={\left ({5\over 4\sqrt 6}\right )}^{3/5}{{\tilde \delta
}^{3/5}\over \Gamma (2/5)}A_1. \label{eq66}
\end{equation}
In the opposite limit of $R \ll R_2$ we get a simple approximate
expression for ${\bf W}(R)$ from (\ref{eq65}) and
(\ref{eq66}) using the asymptotic expressions for the Bessel
functions at large values of their arguments
\begin{equation}
{\bf W}\approx \sqrt {{5\tilde \delta \over 2\pi
\sqrt{24}}}R^{-1/8}\left [A_1\cos \left (z_1 +{\pi \over 20}\right ) + A_2\sin \left (z_1
-{\pi \over 20}\right )\right ]. \label{eq67}
\end{equation}
Equation (\ref{eq67}) tells that in the region $R_{ms} \ll R \ll
R_{2}$ the inclination angle oscillates with the amplitude of
oscillations proportional to $R^{-1/8}$.

\subsubsection{A WKBJ analysis of (\ref{eqn1}) in the limit of
$\tilde \delta \ll 1$}

The asymptotic expression (\ref{eq61}) and (\ref{eq67}) can be
matched together by a WKBJ solution. Indeed, in the limit $\tilde
\delta \ll 1$ the ratio of the quantities $\lambda $ and $b$
entering (\ref{eqn1}), $\tilde \lambda=\lambda/b$, is large for
the range of $R$ such as $z$ and $z_1$ defined in equations
(\ref{eq59}) and (\ref{eq65}), respectively, are large. Applying
the standard WKBJ scheme we find the solution in this region in
the form
\begin{equation}
{\bf W}\approx {C_1\over (\lambda b )^{1/4}}\cos
\left (\int^{R}_{R_{ms}}\sqrt {\tilde \lambda } dR +\phi_{WKBJ}\right ),
\label{eq68}
\end{equation}
where the constants $C_1$ and $\phi_{WKBJ}$ must be chosen in such
a way  that the solution (\ref{eq61}) is matched in the
corresponding asymptotic limit. It is easy to see that in order to
have such a matching we should require that
\begin{equation}
\phi_{WKBJ}= -{\pi \over 2} {1-\epsilon \over 1-2\epsilon },
\label{eq68a}
\end{equation}
and
\begin{equation}
C_1=6^{1/4}\sqrt {{1-2\epsilon \over \pi K_1 U^{\tau}}}C,
\label{eq69}
\end{equation}
where it is assumed that $K_1$ and $U^{\tau}$ are evaluated at
$R=R_{ms}$ and we use the facts that $R_{ms}=6$ and that $D\approx
x^2/72$ close to $R_{ms}$.

In the limit $R\rightarrow \infty $ we can set the values of
$\lambda $ and $b$ in front of the cosine in (\ref{eq68})  equal
to their Newtonian values. The integral in (\ref{eq68}) can be
represented as $I(R)\equiv\int^{R}_{R_{ms}}\sqrt {\tilde \lambda
}dR =I-\int^{\infty}_{R} \sqrt {\tilde \lambda }  dR $, where
$I=\int^{\infty}_{R_{ms}}  \sqrt {\tilde \lambda } dR $. Taking
into account that the Newtonian value of $\tilde \lambda
=24{\tilde \delta }^{-2}R^{-9/4}$ we have $\int^{\infty}_{R}\sqrt
{\tilde \lambda } dR  \approx {8\sqrt 6\over 5}{\tilde \delta
}^{-1}R^{-5/4}$, and, accordingly,
\begin{equation}
{\bf W}\approx C_1 {{\tilde \delta }^{1/2}\over 24^{1/4}}\cos
\left ({8\sqrt 6\over 5}{\tilde \delta }^{-1}R^{-5/4} - I-\phi_{WKBJ}\right ).
\label{eq70}
\end{equation}
The expression (\ref{eq70}) can be matched to the asymptotic
expression (\ref{eq67}) provided that the constant $A_1$ and $A_2$
are appropriately chosen. A simple calculation gives
\begin{eqnarray}
A_1=\sqrt {2\pi \over 5}C_1 \cos \left.\left (I+\phi_{WKBJ}-{\pi \over
20}\right )\right /\cos {\pi \over 10},\nonumber\\ \quad A_2=\sqrt {2\pi \over 5}C_1 \sin
\left.\left (I+\phi_{WKBJ}+{\pi \over 20}\right )\right /\cos {\pi \over 10}. \label{eq71}
\end{eqnarray}

Equations (\ref{eq58}), (\ref{eq65}) and (\ref{eq68}) provide
expressions for an approximate shape of a low viscosity twisted
disc in the whole allowed range of $R$, $R_{ms} < R < \infty $.
Note that although we derive these equations assuming that $a > 0$
they are formally valid for negative values of $a$ as well.

Equations (\ref{eq60}), (\ref{eq66}), (\ref{eq69}) and (\ref{eq71})
allow us to relate the  asymptotic value at large radii, ${\bf
W}_{\infty}$, to the value of ${\bf W}$ at the last stable orbit,
${\bf W}_0$, as
\begin{equation}{\bf W}_\infty = C_{tot}(\tilde
\delta) {\bf W}_0, \label{eq71a} \end{equation} where an explicit
form for $C_{tot}(\tilde \delta )$ follows from these equations.
It is important to note that $C_{tot}(\tilde \delta )\propto \cos
(I+\phi_{WKBJ}-{\pi \over 20})$ as follows from equations
(\ref{eq66}) and (\ref{eq71}). Thus, for values of $\tilde \delta
$ such that $\cos (I+\phi_{WKBJ}-{\pi \over 20})=0$, ${\bf
W}_\infty =0$ while ${\bf W}_0\ne 0$. This describes a peculiar
"resonant" solution to equation (\ref{eqn1}), where a regular at
the last stable orbit solution is matched precisely to the
solution of (\ref{eq64}) proportional to $A_2$ while $A_1=0$, see
equation (\ref{eq65}). From equation (\ref{eqn2}) it follows that
the integral $I$ in the cosine can be written in the form
$I=\tilde \delta^{-1} \tilde I$, where $\tilde I$ does not depend on
$\tilde \delta $, we have $\tilde I_{Th}=1.01$ and $\tilde
I_{ff}=0.69$. This allows us to write the condition for obtaining
the resonant solutions as a condition that the parameter $\tilde
\delta $ has discrete values
\begin{equation}
\tilde \delta_k = {\tilde I\over {\pi \over 2}{\left ({11\over 10}+
{1-\epsilon \over 1 -2\epsilon}+2k\right)}}, \label{eq72}
\end{equation}
where $k$ is an integer. Thus, a stationary twisted disc described
by such a solution has its inclination angle going to zero with
radius. Note, however, that inclusion of effects determined by the
viscosity coefficient $\alpha $ leads to a partial suppression of
this resonance. As we see later numerical results show that when
the resonant condition (\ref{eq72}) is fulfilled and  a small but
nonzero value of $\alpha $ is considered the value of $C_{tot}$ at
the resonant values of $\tilde \delta $ is much smaller than its
neighbouring values instead of being equal precisely to zero.

Neglecting the possibility of the resonances we can roughly
estimate the dependence of $C_{tot}$ on $\tilde \delta $ using
very simple arguments. Indeed, from equation (\ref{eq67}) it
follows that the inclination angle scales with $R$ as ${\bf
W}\propto R^{-1/8}$ for radii $R_{ms} < R < R_2$, where $R_2$ is
given by equation (\ref{eq63a}) while when $R > R_2$ it stays
approximately constant. Thus, we can roughly estimate ${\bf
W}(R\sim R_{ms}) \sim R_{2}^{1/8}{\bf W}_{\infty}={\tilde
\delta}^{-1/10} {\bf W}_{\infty}$. This estimate does not take
into account the growth of the amplitude of the oscillations close
to the last stable orbit, which is described by equation
(\ref{eq62}). Using this equation and assuming, for simplicity,
that the numerical factor $6{\tilde
\chi}^{(1-\epsilon)/2(1-2\epsilon)}\sim 1$ there, we estimate
${\bf W}_0/{\bf W}(R\sim R_{ms})\sim {\tilde
\delta}^{-(1-\epsilon)/(1-2\epsilon)}$. Combining both estimates
we get
\begin{equation}
C_{tot}\sim {\tilde \delta}^{(1-\epsilon)/(1-2\epsilon)+1/10}.
\label{eq73}
\end{equation}

\subsubsection{Restrictions}

We derive our equations assuming that the inclination angle $\beta
$ is small. Also, we suppose that the NT model of the flat
background disc with a constant value of the parameter $\alpha $
is valid. These assumptions put rather severe limitations on
validity of our analysis, especially in the case of the low
viscosity twisted disc considered in Section 4.2.

Since in this case $\gamma \approx 0$ and $\beta \approx {\bf W}$
the condition $\beta < 1$ everywhere in the region $R > R_{ms}$
leads to the condition ${\bf W}_0 < 1$. Equation (\ref{eq73})
tells that in this case the asymptotic value of the inclination
angle at spacial infinity should be quite small
\begin{equation}
{\bf W}_\infty < {\bf W}_{crit}={\tilde
\delta}^{(1-\epsilon)/(1-2\epsilon)+1/10}.\label{eq74}
\end{equation}

Another somewhat more stringent constraint stems from
the fact that the shear velocities induced in the disc by the disc
twist should not be too large. When considering the oscillations
of the disc inclination angle at scales $\sim R_2$ II supposed
that a characteristic amplitude of the shear velocities should not
exceed the sound speed, $c_{s}$. In the opposite case the shearing
instability could make the disc more turbulent, thus increasing an
effective value of $\alpha $. Another possibility is the presence
of shocks in the disc (e.g. \citet{FB08}), which may
heat up the disc thus effectively increasing the relative disc
thickness~$\delta $.

II obtained a condition that the relativistic oscillations do not
induce such velocity amplitudes at the scale $\sim R_2$ provided
that ${\bf W}_\infty < 2\delta_{*}^{4/5}$. Note, however, that our
numerical results suggest that viscosity effects suppress the
oscillation amplitude, and, accordingly, the velocity amplitude
more efficiently than it was assumed in a simplified analysis of
II. Therefore, the constraint of II may to be too restrictive for
more realistic models taking into account a small but non-zero
value of $\alpha$, see the next Section.

In order to estimate a characteristic amplitude of the shear
velocities induced in the disc we assume that it is of the order
of $v_{sh}=h{\bf B}$, where ${\bf B}$ is defined in equation
(\ref{eq41}) and we remind that it can be considered as a real
quantity as far as equation (\ref{eqn1}) is concerned. From
equation (\ref{eq50}) it follows that ${\bf B}\approx
U^{\varphi}{\bf W}^{\prime}$ when $\dot {\bf B}$ and $\alpha$ are
equal to zero there and $\kappa \approx 0$, which is appropriate
for the radii close to $R_{ms}$. On the other hand, from the
results of Section 2.4 it follows that $c_s=U^{\varphi}h/R$.
Introducing the variable $\tilde v= v_{sh}/c_s$ we have $\tilde
v=R{\bf W}^{\prime}\approx 6{\bf W}^{\prime}$. From equation
(\ref{eq61}) it follows that when $x < 1$ but $z > 1$ we have
${\bf W}\sim C\sqrt{1\over xz}z^{\prime}\sim
C{\chi}^{1/4}x^{-(1+\epsilon)}$ and from equation (\ref{eq60}) it
follows that $C\sim \chi^{-(1-2\epsilon)/4}{\bf W}_{0}$, and,
therefore,
\begin{equation} \tilde v \sim 6{\bf W}_{0}\chi^{\epsilon
/2}x^{-(1+\epsilon)}. \label{eq75}
\end{equation}
Since $\chi=\tilde \chi {\tilde \delta }^{-2}$ and the factor
${\tilde \chi}^{\epsilon /2}\sim 1$ we can roughly estimate that
$\chi \sim {\tilde \delta}^{-2}$ in (\ref{eq75}). We see that the
condition $\tilde v < 1$ is fulfilled when
\begin{equation}
x > x_{crit}=(6\tilde {\bf W}_{0})^{1/(1+\epsilon )}{\tilde
\delta}^{-\epsilon/(1+\epsilon)}. \label{eq76}
\end{equation}
According to our criterion our analysis is valid close to $R_{ms}$
provided that we demand that $x_{crit} < 1$, at least. This leads
to inequality ${\bf W}_{0} <{\tilde \delta}^{\epsilon}/6$, which
can be reformulated in terms of ${\bf W}_\infty$ using equation
(\ref{eq73})
\begin{equation}
{\bf W}_\infty < {\bf W}_{crit}{\tilde \delta}^{\epsilon}/6,
\label{eq77}
\end{equation}
where ${\bf W}_{crit}$ is defined in (\ref{eq74}). We see that
this inequality is stronger than the inequality  (\ref{eq74})
although the difference is not quite significant since the
parameter $\epsilon $ is typically rather small.

\subsection{Numerical results}

We integrate numerically equation (\ref{eq52}) starting from the
last stable orbit. Similar to the case of almost inviscid twisted
disc considered above there is a partial solution regular at the
last stable orbit and we use this solution to specify the boundary
condition at $R_{ms}$.

We consider below the case of Thomson opacity only. The case
of free-free opacity is quite similar.

\begin{figure}
\begin{center}
\vspace{1cm}
\includegraphics[width=16cm,angle=0]{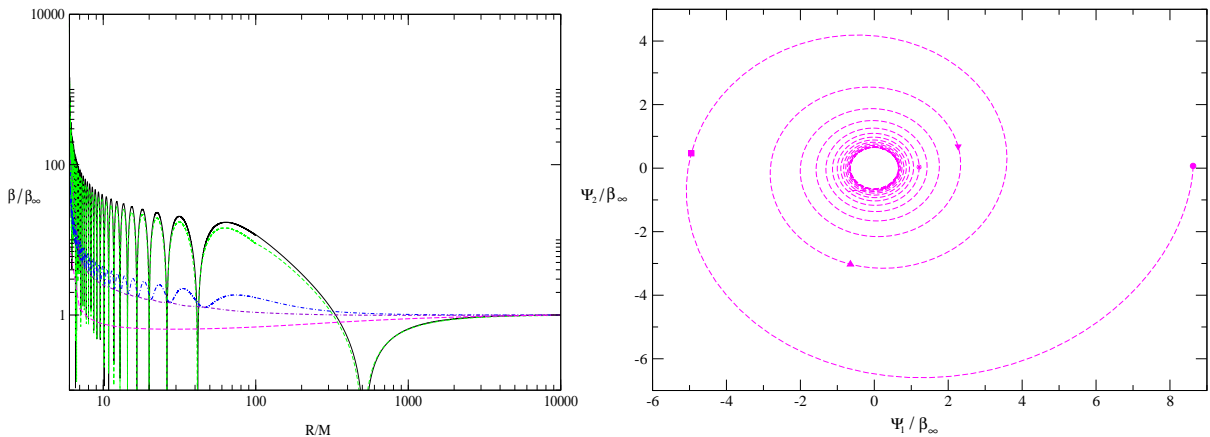}
\end{center}
\caption{ {\it Left panel:} The dependence of $\beta$ on the
radial coordinate $R$ is shown. Since the problem is a linear one
we set hereafter the value of the inclination angle at large
radii, $\beta_{\infty}$, equal to unity in all Figs. The parameter
$\tilde \delta =10^{-2}$. The solid curve is obtained by setting
$\alpha=0 $ in (\ref{eq52}). The short dashed curve represents our
analytical results discussed above. We use the expression for
$\beta $ given by equation (\ref{eq68}) in an intermediate region
of $R$, when $R-R_{ms}\ll 1$ the expression (\ref{eq58}) is used
and when $R \sim R_{2}$ or larger we use equation (\ref{eq65}).
The dot-dashed, dot-double dashed, long dashed curves correspond
to $\alpha =10^{-4}$, $10^{-3}$ and $10^{-2}$, respectively. {\it
Right panel:} The result of integration of equation (\ref{eq52})
presented in parametric form. $\Psi_{1}=Re{\bf W}=\beta \cos
\gamma$ and  $\Psi_{2}=Im{\bf W}=\beta \sin \gamma$, $\tilde
\delta =10^{-2}$ and $\alpha=10^{-2}$ for this plot. The circle,
square, triangle (pointing up), triangle (pointing down) and the
star correspond to $R=6$, $6.1$, $6.2$, $6.3$ and $7$,
respectively.} \label{figg1}
\end{figure}

\subsubsection{Dependencies of $\beta $ and $\gamma $ on $R$}

\begin{figure}
\begin{center}
\vspace{1cm}
\includegraphics[width=9cm,angle=0]{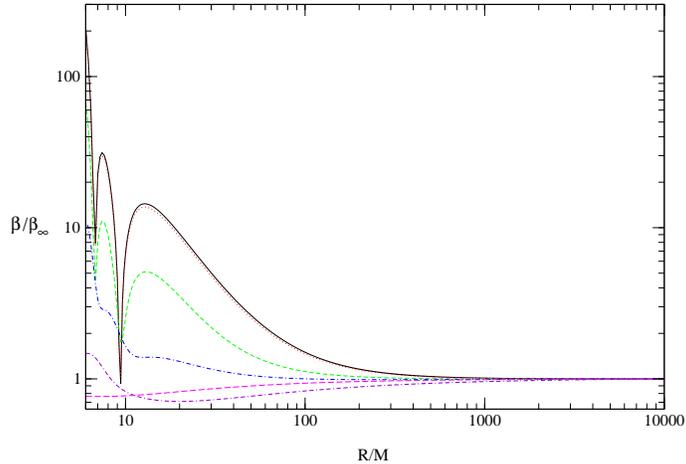}
\end{center}
\caption{The same as on the left panel in Fig. \ref{figg1} but for $\tilde
\delta =10^{-1}$. The solid, dotted, short dashed, dot-dashed, dot-double dashed and long dashed curves
correspond to $\alpha = 0, 10^{-4}$, $10^{-3}$, $10^{-2}$, $10^{-1}$ and $1$, respectively.}
\label{figg2}
\end{figure}

At first let us discuss the dependency of $\beta $ on $R$ for the
case of relatively small viscosity parameter $\alpha \le 10^{-2}$
and $a > 0$. This case is represented in Figs. \ref{figg1} (left
panel) and \ref{figg2}, for $\tilde \delta =10^{-2}$ and $0.1$,
respectively. \footnote{Note that we show the ratios $\beta(R)/\beta_\infty$
 in these Figs, using the fact that our problem is a linear one.
This ratio can be arbitrary large provided that $\beta_\infty$ is sufficiently small. }
As seen from these Figs. the oscillations of the
inclination angle are quite prominent for the case when $\alpha
=0$ in equation (\ref{eq52}) as well as for the case of quite
small $\alpha=10^{-4}$. Also, it is seen that our analytic theory
described above gives the curve (the left panel in Fig.
\ref{figg1}), which is in a quite good agreement with the
numerical one calculated for $\alpha=0$. Effects of viscosity tend
to smooth out the oscillations, which disappear for the curves
with $\alpha=10^{-3}$ and $10^{-2}$ on the left panel in Fig.
\ref{figg1}. The similar tendency is seen in Fig. \ref{figg2}
where, additionally, the case of relatively large $\alpha>10^{-2}$
is shown as well. It is clear from these Figs. that when $\alpha$
is small the inclination angle grows with decrease of $R$ either
on average or monotonically. Thus, in these cases the disc does
not align with the black hole equatorial plane and the
Bardeen-Petterson effect is absent. Note that the smooth curves do
not have sharp gradients of the inclination angle and,
accordingly, they correspond to solutions, where the shear
velocities induced in the disc are relatively small. Therefore,
these solutions may be physically realistic even for relatively
large values of the inclination angle at large radii,
$\beta_{\infty}$.

Instead of showing dependencies of the second Euler angle $\gamma
$ on $R$ we plot curves corresponding to our solutions on the
plane $(\Psi_{1}=\beta \cos \gamma ,\Psi_{2}=\beta \sin \gamma )$,
which depend parametrically on $R$. Since the angle $\gamma $ does
not change significantly with $R$ for very small values of $\alpha
$ and the corresponding curves are close to straight lines, the
case of small viscosity parameters is represented by only one
curve corresponding to the solution with $\alpha=10^{-2}$, see the
right panel in Fig. \ref{figg1}. One can see from this Fig. that
this curve has a form of a tight spiral. In this case $\beta_0\sim
8$ and $\gamma=0$, so the curve starts at $R=R_{ms}$ when $\Psi_1
\sim 8 $ and $\Psi_2=0$. It spirals clockwise with increase of
$R$, i.e. in the direction opposite to the orbital motion and
black hole rotation, which are counterclockwise.

\begin{figure}
\begin{center}
\vspace{1cm}
\includegraphics[width=16cm,angle=0]{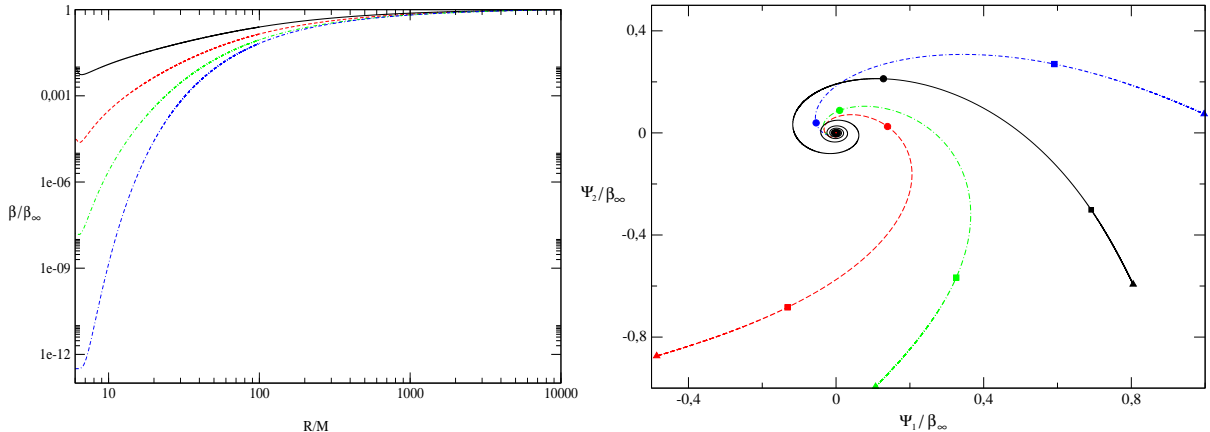}
\end{center}
\caption{Same as in Fig. \ref{figg1} but for larger values of
$\alpha$. The solid, dashed, dot-dashed and dot-double dashed
curves are plotted for $\alpha =0.05$, $0.1$, $0.2$ and $1$,
respectively. Curves of the same style correspond to the same
$\alpha$ on both panels. Different symbols on the right panel
correspond to different values of $R$, $R=100$, $10^3$ and $10^4$
for the circles, squares and triangles, respectively. }
\label{figg3}
\end{figure}

The dependence of $\beta $ on $R$ as well as the solutions in the
parametric form for $\tilde \delta =10^{-2}$ and larger values of
$\alpha =0.05$, $0.1$, $0.2$ and $1$ are shown in Fig.
\ref{figg3}. From the left panel in Fig. \ref{figg3} one can see
that in this case the inclination angle monotonically decrease in
practically the whole range of $R$ while there is a small increase
of the angle close to $R_{ms}$.  In general, the ratio of
$\beta_{0}/\beta_{\infty}$ is quite small and this may be
interpreted as manifestation of the Bardeen-Petterson effect.
Contrary to the case of the low viscosity discussed above the
initial value of $\beta_0$ is smaller than one, thus for the
curves on the right panel of Fig. \ref{figg3} $\beta =\sqrt
{\Psi_1^2+\Psi_2^2}$ increases its value when the curves spiral
from $R_{ms}$ to larger $R$. The direction of spiralling with the
increase of $R$ is also clockwise similar to what is shown on
right panel of Fig. \ref{figg1}.

However, returning to Fig. \ref{figg2} we see that for
$\tilde\delta=10^{-1}$ the Bardeen-Petterson effect is absent even
when $\alpha\sim 1$ in the sense that the ratio
$\beta_{0}/\beta_{\infty}$ remains to be of order of unity.

Thus, in the case of $a > 0$ the Bardeen-Petterson effect occurs
only when the viscosity parameter $\alpha $ is sufficiently large
and $\tilde\delta$ is sufficiently small.

\begin{figure}
\begin{center}
\vspace{1cm}
\includegraphics[width=16cm,angle=0]{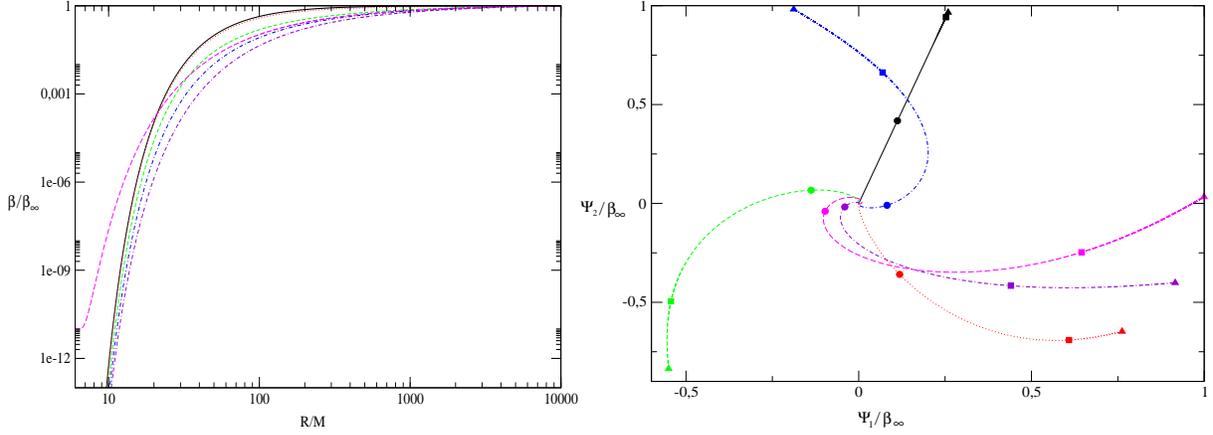}
\end{center}
\caption{The same as in Fig. \ref{figg1} but for the case of the
black hole rotating in the opposite sense with respect to the disc
rotation, $a < 0$. The solid, dotted, short-dashed, dot-dashed,
dot-double dashed and the long-dashed curves correspond to
$\alpha=0$, $0.01$, $0.05$, $0.1$, $0.2$ and $1$, respectively.
Curves of the same style correspond to the same $\alpha$ on both
panels and positions of different symbols on the right panel
correspond to the same $R$ as in Fig. \ref{figg3}.} \label{figg4}
\end{figure}

The case of $a < 0 $ is shown in Fig. \ref{figg4}, for
$\tilde \delta =10^{-2}$ and a range of values of $\alpha $. As is
seen from this Fig. contrary to the previous case the disc always
aligns with the black hole. This is valid even in the case of
setting $\alpha =0$ in (\ref{eq52}), which is described by the
solid curve in Fig. \ref{figg4}. The curves with small values of
$\alpha $ are close to this limiting curve. It is interesting to
point out that as seen from the right panel in Fig. \ref{figg4}
the direction of spiralling when $R$ runs from smaller to larger
radii is opposite to the case of $a >0$. Now the curves spirals in
the direction of the orbital motion but, again, opposite to the
direction of the black hole rotation. Thus, the direction of
spiralling is always opposite to the black hole rotation. This may
have some observational consequences.

\subsubsection{An analysis of behaviour of the solutions for the case $a > 0$ on the parameter
plane ($\alpha,\tilde\delta$)}

As we discuss above our equation contains two independent
parameters - $\alpha $ and $\tilde \delta $. It is of interest to
describe how the most important characteristics of the solutions
such as the ratio $\beta_{0}/\beta_{\infty}$ depend on them.

\begin{figure}
\begin{center}
\vspace{1cm}
\includegraphics[width=9cm,angle=0]{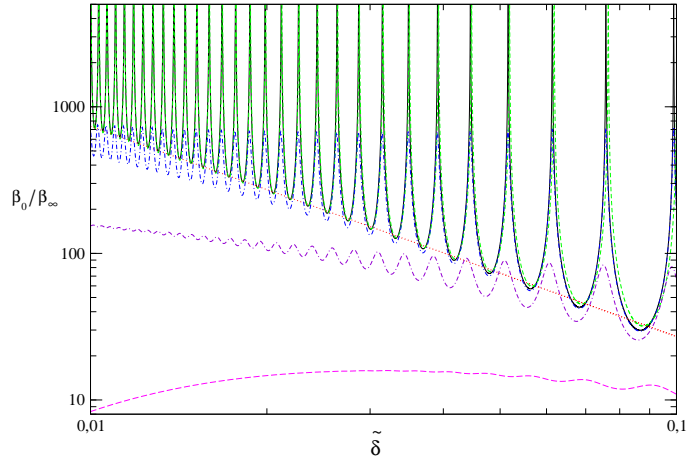}
\end{center}
\caption{The dependence of the ratio of the
inclination angle at the last stable orbit to the inclination
angle at a large radius, $\beta_{0}/\beta_{\infty}$, on the
parameter $\tilde \delta$. The solid curve is calculated for
$\alpha$ set to zero in  (\ref{eq52}), the short dashed curve
represents $C_{tot}^{-1}(\tilde \delta )$, where $C_{tot}$ is
defined in equation (\ref{eq71a}) while the dotted curve
represents our simple estimate (\ref{eq73}), which does not take
into account the effect of the resonances. The dot-dashed,
dot-double dashed and the long dashed curves are for
$\alpha=10^{-4}$, $10^{-3}$ and $10^{-2}$, respectively.}
\label{figg5}
\end{figure}

In Fig. \ref{figg5} we show the dependence of this ratio on
$\tilde \delta $ for various sufficiently small values of
$\alpha$. One can see that the curve corresponding to $\alpha =0$
experiences a quite non-monotonic behaviour with series of peaks at
some successive values of $\tilde \delta $. Since the analytical
curve based on our WKBJ theory developed above is in excellent
agreement with the numerical one we can identify these peaks with
the resonant solutions of (\ref{eq52}) with $\alpha=0$ for which
the value of the inclination angle tends to zero when
$R\rightarrow \infty$. The positions of these peaks are,
accordingly, given by equation (\ref{eq72}). Note also that our
simple estimate (\ref{eq73}) of the dependence of
$\beta_{0}/\beta_{\infty}$ on $\tilde \delta $ provides a very
good approximation to the numerical curve when the resonance peaks
are not taken into account.

From the analytical theory it follows that when $\alpha=0$
formally the amplitude of peaks in (\ref{figg5}) is infinite. When
a small value of $\alpha $ is taken into account the peak's
amplitudes get finite values decreasing with increase of $\alpha $
and/or $\tilde \delta $. For a relatively large value of $\alpha
=10^{-2}$ the peaks practically disappear and the corresponding
curve strongly deviates from the theoretical one. In the case the
values of $\beta_{0}/\beta_{\infty}\sim 10$ in the whole range of
considered values of $\tilde \delta $.

\begin{figure}
\begin{center}
\vspace{1cm}
\includegraphics[width=9cm,angle=0]{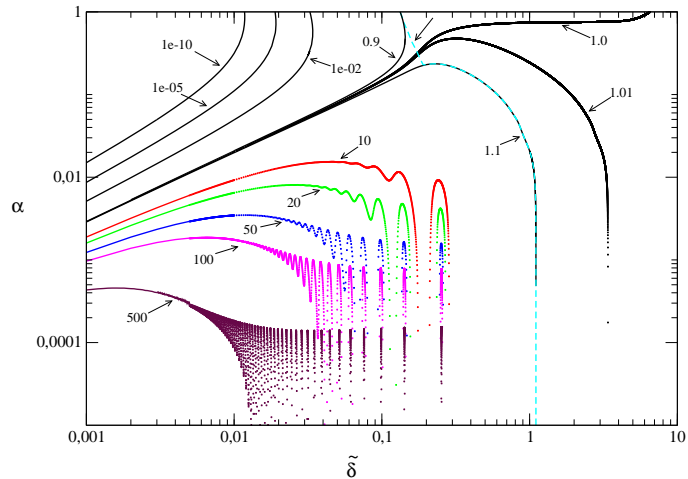}
\end{center}
\caption{Levels of constant ratios of $\beta_{0}/\beta_{\infty}$
on the plane $(\tilde \delta$, $\alpha )$ of the parameters of the
problem. Arrows show the values of $\beta_{0}/\beta_{\infty}$ for
each curve. The dashed curve on the right hand side of the plot
separates a region on the parameter plane, where a change of
$\beta $ with $R$ is larger than 10 per cent of $\beta_{\infty}$
(to the left of this curve) from the region, where the black hole
influence on the disc twist is relatively small and the change is
less than 10 per cent (to the right of the curve).} \label{figg6}
\end{figure}

\begin{figure}
\begin{center}
\vspace{1cm}
\includegraphics[width=9cm,angle=0]{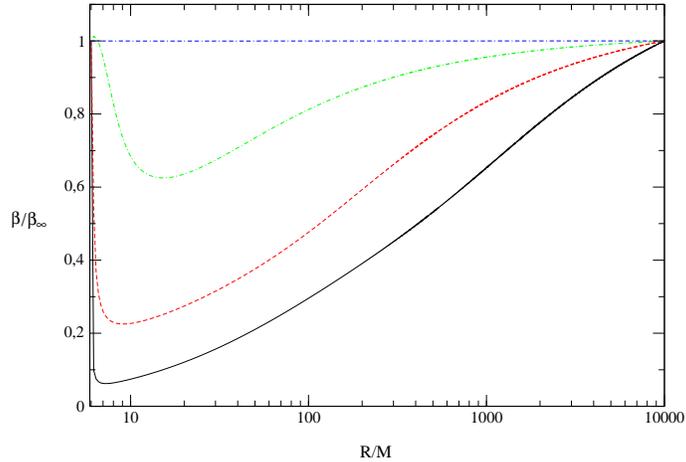}
\end{center}
\caption{Dependencies of $\beta $ on $R$ along the curve in Fig.
\ref{figg6}, where  $\beta_{0}/\beta_{\infty}=1$. The  values of
$\tilde \delta $  are equal to $10^{-3}$, $10^{-2}$, $10^{-1}$ and
$1$ for the solid, dashed, dot-dashed and dot-double dashed
curves, respectively.} \label{figg7}
\end{figure}

In Fig. \ref{figg6} we show the curves of constant values of the
ratio of $\beta_{0}/\beta_{\infty}$ on the plane $(\tilde \delta$,
$\alpha )$ of the parameters entering equation (\ref{eq52}), for
the case of $a > 0$ and the value of $\tilde \delta $ in the range
$10^{-3} < \tilde \delta < 10$. When this ratio is small the disc
aligns with the black hole equatorial plane at small radii. One
can see that the area of small values of
$\beta_{0}/\beta_{\infty}$ is situated at the upper left part of
the plot. The opposite case, where $\beta_{0}/\beta_{\infty} > 1$
corresponds to a situation when the Bardeen-Petterson effect is
absent. For the considered range of $\tilde \delta $ the area in
the plot, where $\beta_{0}/\beta_{\infty} > 1$ is larger than the
area corresponding to the small values of this ratio. The curves
corresponding to the levels of  $\beta_{0}/\beta_{\infty}\ge 10$
exhibit oscillatory behaviour. This effect is most probably
determined by the resonant behaviour of $\beta_{0}/\beta_{\infty}$
at small values of $\alpha $ considered above.

The dashed curve on the right hand side of the plot separates the
regions on the parameter plane where the quantity $|\beta (R)
/\beta_{\infty}-1|$ is smaller/larger than $0.1$, for all values
of $R$. The latter condition is satisfied to the left of this
curve, where the disc twist induced by the black hole is
sufficiently large. Thus, the black hole significantly influences
on the disc only when the value of $\tilde \delta $ is
sufficiently small, $\tilde \delta < 0.1-1$ depending on value of
$\alpha$.

There is a curve in the plot corresponding to the level of
$\beta_{0}/\beta_{\infty}=1$. The dependencies of $\beta (R)$ with
parameters $\alpha $ and $\tilde \delta $ taken along this curve
are shown in Fig. \ref{figg7}. As is seen from this Fig. when
$\tilde \delta $ is sufficiently  small the angle $\beta $ behaves
in a non-monotonic way, though the curves are smooth and no
oscillations of the inclination angle are observed. Such a
behaviour looks realistic and it may have serious consequences for
the disc structure itself since the inner parts of the disc
strongly irradiate the outer part in such cases. Also, it may have
some interesting outcomes for spectra modelling, etc.

\section{Discussion}

In this Paper we derive dynamical equations describing evolution
and stationary configurations of a fully relativistic thin twisted
disc around a slowly rotating black hole. We assume that the disc
inclination angle $\beta \ll 1$, the black hole rotational
parameter $a$ is small and that the background flat accretion disc
is described by the simplest Novikov-Thorne model with a constant
value of the Shakura-Sunyaev parameter $\alpha $. With these
assumptions being adopted a characteristic evolution time scale of
such a disc is much larger than the dynamical time scale. This
helps to simplify our analysis to a great extent. It is shown that
the final dynamical equations may be formulated as equations for
two dynamical complex variables ${\bf W}$ describing the disc
twist and warp and ${\bf B}$, which describes shear velocities
induced in such a disc. They may be brought to a form similar to
what has been used for description of the classical Newtonian
twisted discs, see equations (\ref{eq50}) and (\ref{eq51}).

Since all our variables are well defined in the black hole
space-time, the results of integration of our dynamical equations
may be directly applied to the problems connected with
observations, e.g. obtaining time-dependant luminosity and
spectrum of a twisted accretion disc, its shape as seen from
large distances, etc..

We analyse stationary configurations of twisted discs in our model
and show that they can be fully described by only two parameters -
the viscosity parameter $\alpha $ and by the parameter $\tilde
\delta =\delta_{*}/\sqrt {|a|}$, where $\delta_{*}$ is a
characteristic opening angle of a background flat accretion disc.

We consider in details the case of a very low viscosity twisted
disc formally setting $\alpha = 0$ in equation (\ref{eq52})
describing the stationary configurations and develop an analytic
approach to the problem of finding of these configurations for the
most interesting case $\tilde \delta \ll 1$ and $a > 0$, which is
in excellent agreement with results of numerical integration of
(\ref{eq52}). We find that the oscillations of the disc
inclination angle found by II in their model of a low viscosity
twisted disc far from the black hole can be in a different regime
in the relativistic region close to the last stable orbit. Namely,
for the background flat disc models having singular profiles with
the surface density being formally zero at $R_{ms}$ as in the gas
pressure dominated NT models the amplitude of disc oscillations
and the radial frequency can grow indefinitely large when $\tilde
\delta \rightarrow 0$. Additionally, in the formal limit
$\alpha=0$ there are "resonant" solutions occurring at some
discrete values of $\tilde \delta$, which have a finite value of
the disc inclination angle at $R=R_{ms}$ and the disc inclination
tending to zero when $R\rightarrow \infty $.

Effects of non-zero viscosity tend to smooth out the relativistic
oscillations to an extent more prominent than was
obtained in an oversimplified model of II. This may explain the
results of \citet{NP00} who didn't find prominent
oscillations of the disc inclination angle in their SPH
simulations of the Bardeen-Petterson effect. When viscosity
parameter $\alpha $ gets sufficiently large (say, $\alpha \sim
10^{-3}-10^{-2}$ for $\tilde \delta =10^{-2}$) the disc
oscillations are not observed. Instead of it there is a growth of
the inclination angle towards the last stable orbit, which is
opposite to what is expected in the standard picture of the
Bardeen-Petterson effect. When $\alpha $ gets even larger (say,
$10^{-1}$ for $\tilde \delta =10^{-2}$) the inclination angle
decreases towards $R_{ms}$ as in the standard picture.

In general, there are three possible qualitatively different
shapes of the twisted disc depending on values of the parameters
of the problem.

1) When $\alpha $ is very small the disc inclination angle
oscillates, with the oscillation amplitude growing dramatically
towards $R_{ms}$. When the disc inclination angle at large radii
is not very small such a regime is probably unphysical. As we
discuss in the text in such a case the shear velocities induced in
the disc can easily exceed the speed of sound. This may either
make the disc more turbulent thus effectively increasing the value
of $\alpha $, which may, in its turn, damp the oscillations. On
the other hand this can lead to presence of shocks in the disc
close to $R_{ms}$. The energy dissipated in the shocks may make
the disc thicker. This opens up a possibility of explaining of the
presence of more compatible with observations thick discs close to
$R_{ms}$ by such a mechanism.

2) When $\alpha $ is moderately small the oscillations are absent.
However, the Bardeen-Petterson effect is absent as well. In such a
regime the disc inclination can either monotonically grow towards
the last stable orbit or exhibit a non-monotonic behaviour
decreasing with decrease of $R$ at large radii and increasing at
smaller radii, see Figs. \ref{figg1} and \ref{figg7}. Since the
corresponding curves are quite smooth we do not expect large shear
velocities, which are proportional to the gradient of ${\bf W}$.
Therefore, such a regime may be physically allowed. This can lead
to interesting possibilities of modifying the disc structure by
strong irradiation of outer parts of the disc by inner parts, can
be very important for modelling of spectra of the discs, etc..

3) Finally,  when $\alpha $ is sufficiently large the
Bardeen-Petterson effect is observed. The region of the parameter
plane, where such effect is possible is shown in Fig.
\ref{figg6}, for the considered range $10^{-3} < \tilde \delta <
10$.

Although we consider the rotational parameter of the black hole as
a small parameter of the problem we believe that the outlined
picture remains qualitatively correct for moderate values of the
rotational parameter as well, say when $a\sim 0.5 $.

We discuss possible generalisations of our results. 
At first one can solve time-dependant
equations (\ref{eq50}) and (\ref{eq51}) to clarify dynamics of the
twisted discs in the relativistic regime. This will be done in a
separate publication. Secondly, one can try to generalise the
results to the case of large values of the rotational parameter,
$a\sim 1$. We expect that time-dependent equations will be rather
cumbersome since in this case the evolution time scale will be of
the order of the dynamical one and many terms neglected in our
analysis must be taken into account. The generalisation to the
stationary case looks relatively straightforward. However, it may
happen that our twisted coordinate system based on the cylindrical
coordinate system is less convenient when $a\sim 1$ than a twisted
coordinate system based on the spherical coordinates similar to
what was introduced by \citet{Og99} for the classical Newtonian
twisted discs. One may also try to include (at least, at some
phenomenological level) the effects of back reaction on the
background flat disc model determined by the large shear
velocities induced in the twisted disc, and the disc
self-irradiation. Additionally, as we discuss in Section 2.4
equations describing the flat disc models used in this paper are
invalid  very close to the last stable orbit, where the radial
drift velocity gets larger than the sound speed, see equation
(\ref{eqmm4}). One may try to use a slim disc model close to
$R_{ms}$ as a background model for our equations to overcome this
difficulty. Finally, one may use the shape of the stationary
configurations calculated in this paper to model different
observed characteristics of particular sources.

\section*{Acknowledgments}

We are grateful to N.I. Shakura for many helpful and fruitful
discussions. We also thank M. A. Abramowicz and A.F. Illarionov
for useful conversations, M. Buck and M. H. Moore for helpful comments.

VZh was supported in part by ''Research and Research/Teaching
staff of Innovative Russia'' for 2009 - 2013 years (State Contract
No. P 2552 on 23 November 2009) and in part by the grant RFBR-09-02-00032.

PBI was supported in part by the Dynasty Foundation, in part by
''Research and Research/Teaching staff of Innovative Russia'' for
2009 - 2013 years (State Contract No. P 1336 on 2 September 2009)
and in part by the grant RFBR-11-02-00244-a.

\begin{appendix}

\section{The one forms and connection coefficients}

\subsection{The forms}
The transformation law (\ref{eq7}) may be considered as a rotation
to another Cartesian coordinate system $(\tau=t, x_1=r\cos \psi, y_1=r\sin
\psi, z_1=\xi)$ with help of the rotational matrix entering
(\ref{eq7}). Let us call this matrix as $\bf A$. The one forms
(\ref{eq9}-\ref{eq12}) are related to the forms associated with the new coordinates
$(\tau, x_1, y_1, z_1)$ as the forms (\ref{eq6}) are
related to the forms associated with the old one $(t, x, y, z)$. Explicitly, the
transformation between the forms (\ref{eq6}) and the forms
associated with the coordinates $(t,x,y,z)$ may be written as
\begin{equation}
\mbox{\boldmath$\bar\omega$}_{Car} ={\bf B} \mbox{\boldmath$\bar\omega$}_c, \label{An1}
\end{equation}
where $\mbox{\boldmath$\bar\omega$} $ are vectors having the corresponding
one forms as their components, e.g. $\mbox{\boldmath$\bar\omega$}_c$ has the
components $(\mbox{\boldmath$\omega$}^{(t)},
\mbox{\boldmath$\omega$}^{(r)},
\mbox{\boldmath$\omega$}^{(\phi)},
\mbox{\boldmath$\omega$}^{(z)})$, etc.. $\bf B$ is a
rotational matrix with components
\begin{equation}
\begin{array}{cccc}
\left ( \begin{array}{cccc}
1 & 0 & 0 & 0 \\
0 &  cos \phi & -\sin \phi & 0 \\
0 & sin \phi & cos \phi & 0  \\
0 & 0 & 0 & 1
\end{array} \right )
\end{array}. \label{An2}
\end{equation}
Analogously, the transformation between the vector containing the
forms (\ref{eq9}-\ref{eq12}), $\mbox{\boldmath$\bar\omega$}_{tw}$, and the
vector containing the forms corresponding to the new Cartesian
system, $\mbox{\boldmath$\bar\omega$}_{Car New}$, may be written as
\begin{equation}
\mbox{\boldmath$\bar\omega$}_{Car New} ={\bf C} \mbox{\boldmath$\bar\omega$}_{tw},
\label{An3}
\end{equation}
where the rotational matrix ${\bf C}$ is obtained from ${\bf B}$
by change of the angle: $\phi \rightarrow \psi $. Accordingly, we
have
\begin{equation}
\mbox{\boldmath$\bar\omega$}_{tw} ={\bf C}^{T}{\bf A} {\bf B} \mbox{\boldmath$\bar\omega$}_c. \label{An4}
\end{equation}
The rotational matrix ${\bf C}^{T}{\bf A} {\bf B}$ has the
components
\begin{equation}
\begin{array}{cccc}
\left ( \begin{array}{cccc}
1 & 0 & 0 & 0 \\
0 &  1 & (\xi/r) \beta \cos\psi   & \beta \sin \psi \\
0 & -(\xi/r) \beta \cos\psi &  1  & \beta \cos \psi \\
0 &  -\beta \sin \psi & -\beta \cos \psi & 1  \\
\end{array} \right )
\end{array}. \label{An5}
\end{equation}

Since the transformation (\ref{An4}) is given by an orthogonal
matrix the transformation between the corresponding  sets of
adjoint basis vectors is the same.

\subsection{The connection coefficients}

The non-trivial connection coefficients are given by the following
expressions:
\begin{equation}
\begin{array}{ll}

\Gamma_{\tau r \tau} = \frac{K_1^\prime}{K_1 K_2}, & \Gamma_{\tau
r \varphi} = a\frac{K_3}{K_2^2}\,\left ( 1 - \frac{1}{2} \left (r
- \xi Z \right ) K_4 \right ),
\\&\\
\Gamma_{\tau r \xi} = - a\frac{K_3}{K_2^2}
\partial_\varphi Z \left ( 1 - \frac{1}{2r} \left (r^2+\xi^2
\right ) K_4
\right ) , &
\Gamma_{\tau \varphi r} = -\Gamma_{\tau r \varphi},
\\&\\
\Gamma_{\tau
\varphi \xi} = a\frac{K_3}{K_2^2} \left ( Z + \frac{\xi}{2r} \left (r-\xi Z
\right ) K_4 \right ), & \Gamma_{\tau \xi \tau} =
\frac{\xi}{r}\frac{K_1^\prime}{K_1 K_2},
\\&\\
\Gamma_{\tau \xi r} = -\Gamma_{\tau r \xi}, & \Gamma_{\tau \xi
\varphi} = -\Gamma_{\tau \varphi \xi},
\\&\\
 \Gamma_{r \varphi \tau} =
\frac{\xi}{r} \frac{1}{K_1}  \partial_\varphi U - \Gamma_{\tau r
\varphi}, &
\Gamma_{r \varphi r} = \frac{\xi}{r} \frac{1}{K_2}
\partial_\varphi W,
\\&\\
\Gamma_{r \varphi \varphi} = \frac{(r
K_2)^\prime}{r K_2^2} - a\xi \frac{K_3}{K_1 K_2} \partial_\varphi
U, & \Gamma_{r \xi \tau} = \frac{U}{K_1} - \Gamma_{\tau r \xi},
\\&\\
\Gamma_{r \xi r} = \frac{W}{K_2} -
\frac{\xi}{r}\frac{K_2^\prime}{K_2^2}, & \Gamma_{r \xi \varphi} =
-ar\frac{K_3}{K_1 K_2} U,
\\&\\
\Gamma_{r \xi \xi} = \frac{K_2^\prime}{K_2^2}, &
\Gamma_{\varphi \xi \tau} = \frac{1}{K_1} \partial_\varphi U -
\Gamma_{\tau \varphi \xi},
\\&\\
\Gamma_{\varphi \xi r} =
\frac{1}{K_2} \partial_\varphi W, & \Gamma_{\varphi \xi \varphi} =
-\frac{\xi}{r}\frac{K_2^\prime}{K_2^2} - ar\frac{K_3}{K_1 K_2}
\partial_\varphi U,
\\&\\
\\
\end{array}
\label{1A1}
\end{equation}
where $K_4 \equiv (K_3/K_1) (K_1/K_3)^\prime$.\\
Other non-zero connection coefficients are obtained with help of
antisymmetry over the first two indices:
$\Gamma_{abc}=-\Gamma_{bac}$.

\section{Perturbed equations of motion}

As we discuss in Section \ref{twist} our subset of equations
describing the disc twist follows from equations (\ref{eq19})
after the procedure outlined in this Section.

The $\tau$-component of  (\ref{eq19}) gives
\begin{multline}
 K\frac{K_2}{K_1} (U^\tau)^2 \dot \rho_1 +
\left ( 2U^\varphi - ar\frac{K_3}{K_2}\frac{(U^\varphi)^2 +
(U^\tau)^2}{U^\tau} \right ) \frac{K_2}{K_1} \rho \dot v^\varphi +
\frac{1}{r} U^\tau U^\varphi \partial_\varphi \rho_1 + \frac{1}{r}
\frac{(U^\varphi)^2 + (U^\tau)^2}{U^\tau} \rho \partial_\varphi
v^\varphi + \\
\partial_r (\rho U^\tau v^r) + \partial_\xi \rho U^\tau v^\xi +
\frac{(rK_1^2 K_2^2)^\prime}{rK_1^2K_2^2} \rho U^\tau v^r +
F_{\nu}^{\tau} = r\partial_\xi \rho (U^\tau)^2 K \frac{K_2}{K_1} U
+ \frac{\xi}{r} \rho U^\tau U^\varphi \partial_\varphi W,
\label{2A1}
\end{multline}
where
$$
K = \left ( 1 - ar\frac{K_3}{K_2} \frac{U^\varphi}{U^\tau} \right
),$$ and the $r$, $\varphi$ and $\xi$-components give,
respectively,
\begin{multline}
 K\frac{K_2}{K_1} U^\tau \dot v^r +
\frac{U^\varphi}{r} \partial_\varphi v^r - \left
[2\frac{K_1^\prime}{K_1 U^\varphi}  + a\frac{K_1}{rK_2 U^\tau}
\left ( \frac{r^2 K_3}{K_1} \right )^\prime \right ] v^\varphi
+{1\over \rho}F^{r}_{\nu} =\\
 W r \frac{\partial_\xi p}{\rho} -
a\xi\frac{K_3^2}{K_1 K_2} \left ( \frac{K_1}{K_3} \right )^\prime
Z U^\tau U^\varphi, \label{2A2}
\end{multline}
\begin{multline}
 K\frac{K_2}{K_1} U^\tau U^\varphi \dot \rho_1 +
\left ( \frac{(U^\varphi)^2+(U^\tau)^2}{U^\tau} -
2ar\frac{K_3}{K_2}U^\varphi \right ) \frac{K_2}{K_1} \rho \dot
v^\varphi + \frac{(U^\varphi)^2}{r} \partial_\varphi \rho_1 + 2
\frac{U^\varphi}{r} \rho \partial_\varphi v^\varphi +\\
\partial_r (\rho U^\varphi v^r) + \partial_\xi\rho U^\varphi v^\xi +
\frac{(r^2 K_1 K_2^3)^\prime}{r^2 K_1 K_2^3} U^\varphi \rho v^r -
a\frac{K_1}{r K_2} \left ( \frac{r^2 K_3}{K_1} \right )^\prime
U^\tau \rho v^r + F^{\varphi}_{\nu} = \\
K\frac{K_2}{K_1} r \partial_\xi\rho U^\tau U^\varphi U + \frac{\xi}{r} \rho
(U^\varphi)^2 \partial_\varphi W, \label{2A3}
\end{multline}
and
\begin{multline}
K\frac{K_2}{K_1}ru^\tau \dot v^\xi + U^\varphi
\partial_\varphi v^\xi + r\frac{\partial_\xi p_1}{\rho} + \xi
\frac{(U^\varphi)^2}{r} \left ( 1 - 2
ar\frac{K_3}{K_2}\frac{U^\tau}{U^\varphi} \right )
\frac{\rho_1}{\rho} +\\
 2\xi U^\varphi v^\varphi \left [
\frac{K_1^\prime}{K_1} - \frac{K_2^\prime}{K_2} - \frac{ar}{2}
\frac{K_3^2}{K_1 K_2} \left ( \frac{K_1}{K_3} \right )^\prime
\left ( \frac{U^\tau}{U^\varphi} + \frac{U^\varphi}{U^\tau} \right
) \right ] +{r\over \rho}F^{\xi}_{\nu} = \\
-\left [ \frac{K_2}{K_1} \partial_\varphi U - 2a\frac{K_3 Z}{K_2}
+ a\frac{\xi^2}{r} \frac{K_3^2 Z}{K_1 K_2} \left ( \frac{K_1}{K_3}
\right )^\prime \right ] r U^\tau U^\varphi + ar^2\frac{K_3}{K_1}
(U^\varphi)^2
\partial_\varphi U. \label{2A4}
\end{multline}
where $F_{\nu}^{i}$ are determined by the presence of viscous
interactions in the disc. Since these interactions lead to a
non-zero value of the drift component, $U^r$, we include  terms
containing $U^r$ in $F_{\nu}^{i}$ excepting the terms proportional
to $U^{r}$ through the dependence of $v^{\xi}$ on $U^{r}$, see
equation (\ref{eq34}) to keep the notation uniform.

Explicitly, we have from (\ref{eq19})
\begin{multline}
F^{\tau}_{\nu} = {U^{\varphi}\over U^{\tau}}(\partial_\xi
T_{\nu}^{\varphi \xi}-rW \partial_\xi T_{\nu}^{r\varphi }) - r
\partial_\xi \rho U^\tau U^r W, \\
F^{\varphi}_{\nu}=(\partial_\xi T_{\nu}^{\varphi \xi}-rW
\partial_\xi T_{\nu}^{r\varphi }) - r \partial_\xi \rho U^\varphi U^r W,
\quad F_{\nu }^{r}=\partial_\xi T_{\nu}^{r\xi}, \label{2A5}
\end{multline}
and
\begin{multline}
F^{\xi}_{\nu}={1\over rK_{1}K_{2}^{3}} \partial_r
(rK_{1}K_{2}^{3}T_{\nu}^{r\xi}) + \partial_\xi T_{\nu}^{\xi \xi} +
{1\over r}\partial_\varphi T_{\nu}^{\varphi \xi} + \\
\partial_\varphi W ( T_{\nu}^{r\varphi} + T^{r\varphi}_{adv}  )
 + a \frac{K_1}{r K_2} \left (\frac {r^2 K_3}
{K_{1}}\right )^{'} \partial_\varphi Z \left (\frac{U^\varphi}{U^\tau}
T_{\nu}^{r\varphi}+ \frac{U^\tau}{U^\varphi} T^{r\varphi}_{adv} \right ),
\label{2A6}
\end{multline}
where $T^{r\varphi}_{adv}=\rho U^{\varphi}U^{r}$.  $T_{\nu}^{ij}$
are components of the viscous stress tensor given by
equations (\ref{eq20}) and (\ref{eq21}) relevant for our purposes:
\begin{multline}
T_{\nu}^{r\xi}=-{\eta \over K_{2}}(\partial_\xi v^{r} +
U^{\varphi}\partial_\varphi W), \\
 T_{\nu}^{\varphi \xi}=-{\eta
\over K_{2}} \left (\partial_\xi v^{\varphi} - 2a{K_{3}\over
K_{2}}U^{\tau} (U^\varphi)^2 Z \right ), \quad T_{\nu}^{r\varphi}=-\eta r
\left ({U^{\varphi}\over rK_{2}}\right )^\prime, \label{2A7}
\end{multline}
where we use equations (\ref{eq20}) and (\ref{eq21}).

Note that we take into account all explicit terms of zero and
first order in $a$ in equations (\ref{2A1}-\ref{2A7}). As we
discuss in the main text, in fact, the only term linear in $a$,
which is important for our purposes, is the gravitomagnetic term -
the second term in the square brackets on the right hand side of
(\ref{2A4}), see equation (\ref{2A12}) below. Other terms may be
important in other studies and are shown for a reference. The
background quantities $U^{\tau}$, $U^{\varphi}$ and $\partial_\xi
p/\rho$ can also be easily developed up to the linear in $a$
order, and, for completeness, we show them below. We have
\begin{equation} U^\varphi = U_S^\varphi -
a\,\frac{(U_S^\varphi)^2 U_S^\tau}{R^{1/2} (R-2)^{1/2}}, \quad
U^\tau = U_S^\tau - a\,\frac{(U_S^\varphi)^3}{R^{1/2}
(R-2)^{1/2}}, \label{2A8}\end{equation} where $U_{S}^{\varphi}$
and $U^{\tau}_{S}$ are given by (\ref{eq23}), and
\begin{equation}
\frac{\partial_\xi p}{\rho} = -\frac{(U_S^\varphi)^2}{r^2}\xi
\left [ 1 - 6a \frac{K_1}{rK_2} U_S^\tau U_S^\varphi \right
]=\frac{(U^\varphi)^2}{r^2} \frac{\Omega_\perp^2}{\Omega^2}\xi,
\label{2A9}\end{equation} where
\begin{equation}
\Omega_\perp = \Omega - \Omega_{LT}, \label{2A10}
\end{equation}
we remind that $\Omega=R^{-3/2}$  and  the frequency of the
Lense-Thirring precession (see e.g. Aliev $\&$ Galtsov 1981, Kato
1990), $\Omega_{LT}$, is given by the expression
\begin{equation}
\Omega_{LT} = a\frac{K_1 K_3}{K_2^2} = 2 a R^{-3} \label{2A11}
\end{equation}

It is important to note that from equation (\ref{2A4}) it follows
that this characteristic frequency determines the disc's evolution
in a certain regime. In order to show it let us temporarily assume
that we consider a pressureless, inviscid disc with gas particles
moving on circular orbits around a black hole. In this case we can
set $p_{1}=\rho_1=v^{\varphi}=0$ in (\ref{2A4}). The last two
terms on the right hand side can be also neglected since they
contain products of two small parameters: $a(h/r)^2$ and
$a(t_{d}/t_{tw})$, respectively. Accordingly, we obtain from
(\ref{2A4})
\begin{equation}
 \partial_\varphi U= \Omega_{LT} Z.
\label{2A12}
\end{equation}
Remembering that $U=\dot Z$ we see that in this case the disc
rings precess with the precession frequency equal to
$\Omega_{LT}$.

\section{The law of conservation of angular momentum}

The space-time of a Schwarzschild black hole is symmetric  with
respect to transformations belonging to the three-dimensional
group of rotations, $O(3)$. Accordingly, equations of motion
contain three conserved quantities, which may be identified with
three components of angular momentum. Conservation of the
components perpendicular to the axis $z$ requires a special
divergent form of equation (\ref{eq38})  (or (\ref{eq51})). In
order to see that this equation has, indeed, such a form let us
represent the term on the right hand side as $-{1\over
r^2K_2^4}{\partial \over \partial r} F$ and the quantity $F$ as
$F=C_1\cos \varphi + C_2 \sin \varphi $, where $C_1$ and $C_2$ are
functions of $r$ and $\tau $. Introducing the complex notation
used above we rewrite (\ref{eq38}) as
\begin{equation}
\Sigma U^{\tau}U^{\varphi}(\dot {\bf W} -i\Omega_{LT}{\bf W})+
{K_1\over K_2}(\Sigma U^{\varphi } U^{r}+\bar T_{\nu}^{r\varphi}
){\bf W}^{\prime } =-{1\over r^2K_2^4}{\partial \over \partial r}
{\bf F}, \label{C1}
\end{equation}
where ${\bf F}= C_1 +iC_2$. We multiply (\ref{C1}) by $r^2 K^2$
and integrate over $r$ from some radius $r_{in}$ to $r_{out} >
r_{in}$. We get
\begin{equation}
\dot {\bf L} -i{\bf T} + {\bf I}={\bf F}(r_{in})-{\bf F}(r_{out}),
\label{C2}
\end{equation}
where
\begin{equation}
{\bf L}= \int^{r_{out}}_{r_{in}}dr r^2K_2^4\Sigma
U^{\tau}U^{\varphi} {\bf W}, \quad {\bf T}=
\int^{r_{out}}_{r_{in}}dr r^2K_2^4\Sigma
U^{\tau}U^{\varphi}\Omega_{LT} {\bf W}, \label{C3}
\end{equation}
and
\begin{equation}
{\bf I}= \int^{r_{out}}_{r_{in}}dr r^2 K_1 K_2^3(\Sigma U^{\varphi
} U^{r}+\bar T_{\nu}^{r\varphi} ){\bf W}^{\prime }. \label{C4}
\end{equation}
It can be shown that $\bf L$ is proportional to a linear
combination of $x$ and $y$-components of the disc angular momentum
within the region $r_{in} < r < r_{out}$, and $\bf T$ is the
corresponding torque term provided by the gravitomagnetic force.

The term ${\bf I}$ can be transformed with help of equation
reflecting the law of conservation of angular momentum for a
background disc model
\begin{equation}
{1\over r^{2}K_1K_2^3}{d\over dr}(r^{2}K_1 K_2^3(\Sigma U^{\varphi
} U^{r}+\bar T_{\nu}^{r\varphi} ))+2qU^{\varphi}=0, \label{C5}
\end{equation}
where $q$ is flux of radiation from the disc surface
\footnote{This equation can either be obtained from equation
(\ref{eq19}) written for a flat disc or can be derived from the
results of \citet{PT74} and RH.}. We integrate (\ref{C4})
by parts and use (\ref{C5}) to get
\begin{equation}
\dot {\bf L} -i{\bf T} +
2\int^{r_{out}}_{r_{in}}drr^2K_1K_2^3qU^{\varphi}{\bf W}={\bf
F}(r_{in})-{\bf F}(r_{out})+{\bf I}(r_{in})-{\bf I}(r_{out}).
\label{C6}
\end{equation}
Equation (\ref{C6}) shows that when the black hole is not rotating
(${\bf T}=0$) a change of the angular momentum content within a
region of the disc is determined by fluxes of angular momentum
through the region inner and outer boundaries and angular momentum
flux carried by radiation emitted from the disc surface. Thus,
equation (\ref{eq38}) has the required divergent form.

\end{appendix}

\label{lastpage}

\end{document}